\documentclass[pra,aps,twocolumn,superscriptaddress,longbibliography, nofootinbib, showpacs]{revtex4-1}

\usepackage[colorlinks=true, citecolor=red, urlcolor=blue ]{hyperref}
\usepackage{graphicx}
\usepackage{bm}
\usepackage{amsmath,amsfonts}
\usepackage{amsthm}
\usepackage{hyperref}
\usepackage{xcolor}
\hypersetup{
    urlcolor=blue,
    citecolor=blue
           }
\usepackage{braket}
\theoremstyle{plain}
\usepackage{color}
\usepackage{amssymb}
\usepackage{amsthm}
\usepackage{amsfonts}
\usepackage{float}
\usepackage{tabularx}
\usepackage{graphicx}
\usepackage[export]{adjustbox}

\usepackage{mathtools}
\usepackage{esvect}
\usepackage{wrapfig}
\usepackage{amsthm}
\usepackage{verbatim}
\usepackage{bbm}
\usepackage[normalem]{ulem}

\usepackage{enumitem}
\usepackage{fmtcount}
\usepackage{booktabs}
\usepackage{csquotes}
\usepackage{epsfig}

\usepackage{tabularx}
\usepackage{graphicx}
\usepackage{amsmath}
\usepackage{braket}
\usepackage{latexsym}
\usepackage{bm}
\usepackage{graphics,epstopdf}
\usepackage{enumitem}
\usepackage{fmtcount}
\usepackage{booktabs}
\usepackage{csquotes}
\usepackage{epsfig}
\usepackage{soul}

\theoremstyle{plain}

\def\bea{\begin{eqnarray}}
\def\eea{\end{eqnarray}}
\def\ba{\begin{array}}
\def\ea{\end{array}}

\def\beq{\begin{equation}}
\def\eeq{\end{equation}}

\usepackage[normalem]{ulem}
\usepackage{float}
\usepackage{graphicx}  
\usepackage{dcolumn}          
\usepackage{amssymb}
\usepackage{appendix}
\usepackage{physics}   
\usepackage{mathtools}
\usepackage{esvect}
\usepackage{wrapfig}
\usepackage{amsthm}
\usepackage{verbatim}
\usepackage{bbm}

\usepackage[mathscr]{euscript}
\def\Tr{\operatorname{Tr}}

\def\({\left(}
\def\){\right)}
\def\[{\left[}
\def\]{\right]}

\newtheorem{theorem}{Theorem}

\begin{document}


\title{Unbounded entanglement-sustaining sequential local quantum state discrimination}


\author{Debarupa Saha}
\affiliation{Harish-Chandra Research Institute,  A CI of Homi Bhabha National
Institute, Chhatnag Road, Jhunsi, Prayagraj - 211019, India}
\author{Priya Ghosh}
\affiliation{Harish-Chandra Research Institute,  A CI of Homi Bhabha National
Institute, Chhatnag Road, Jhunsi, Prayagraj - 211019, India}
\author{Kornikar Sen}
\affiliation{Harish-Chandra Research Institute,  A CI of Homi Bhabha National
Institute, Chhatnag Road, Jhunsi, Prayagraj - 211019, India}
\affiliation{Departamento de Física Teórica, Universidad Complutense, 28040 Madrid, Spain}
\author{Chirag Srivastava}
\affiliation{Laboratoire d’Information Quantique, CP 225, Université libre de Bruxelles (ULB), Av. F. D. Roosevelt 50, 1050 Bruxelles, Belgium}
\affiliation{Institute of Informatics, National Quantum Information Centre, Faculty of Mathematics, Physics and Informatics,University of Gda\'nsk, Wita Stwosza 57, 80-308 Gda\'nsk, Poland}
\author{Ujjwal Sen}
\affiliation{Harish-Chandra Research Institute,  A CI of Homi Bhabha National
Institute, Chhatnag Road, Jhunsi, Prayagraj - 211019, India}




\begin{abstract}

Two pure orthogonal quantum states can be perfectly distinguished by sequential local action of multiple pairs of parties. 
However, this process typically leads to the complete dissolution of entanglement in the states being discriminated.  
We propose a protocol that allows an arbitrary number of pairs of parties to distinguish between any two orthogonal, entangled, two-qubit pure states using local quantum operations and classical communication, with a success probability greater than that of random guessing, while ensuring that at each step, the individual ensemble states retain a finite amount of entanglement.  Our protocol employs the minimum-error state discrimination approach. For demonstrating the retention of entanglement in the ensemble states at each step, we use logarithmic negativity as well as the concept of entanglement witnessing. 
For a large family of sets of the two states, the success probability of discrimination can be as close as required to unity, while sustaining a finite amount of entanglement in each step.
\end{abstract}

\maketitle

\section{Introduction}
In a quantum information processing task, perfect retrieval of information encoded in quantum states requires perfect identification of the state. However, the no-cloning theorem ~\cite{nc2, nc1} states that unknown quantum states cannot be exactly cloned and, as a result, cannot be perfectly identified or distinguished. This leads to the ongoing challenge of finding optimal measurement strategies to enhance the distinguishability of quantum states and, consequently, improve information extraction. Any quantum state discrimination (QSD) protocol (see reviews ~\cite{Re1, Re2, Re3, Re4, Re5, Re6}) inherently addresses this problem. The optimal strategy often depends on the chosen figure of merit. If the goal is to achieve the maximum confidence level in discrimination, the protocol is referred to as unambiguous state discrimination ~\cite{U2, U3, U1, U4, U5, U7, U6, U8, U9, U11, U12, U13, U14, U15, U10, U16, U17, U18}. In our work, however, we consider the minimum-error state discrimination (MESD) ~\cite{Hels, Me1, Me2, Me3, Me4, Me5, Me6, Me7, Me8, Me10, Me9, Me12, Me13, Me14, Me15, Me16}, where the figure of merit is the average success probability {of correctly discriminating the states chosen from a priorly known ensemble.}

Beyond information processing, QSD is also fundamental to quantum theory. For instance, it provides an operational meaning to conditional mutual entropy ~\cite{ME} and aids in witnessing the dimension of quantum systems ~\cite{DW}. QSD has applications in various domains, including random access codes ~\cite{Ambainis99, Nayak99, Ambainis02, Rac1, Pereira23, Rac2}, metrology ~\cite{Vittorio04, Giovannetti06, MR1, MR0, MR2, MR3, MR4, MR5}, channel discrimination ~\cite{Raginsky01, CD1, Chiribella08, Jencova14, Jencova16, Caiaffa18, Wilde20}, and dense coding ~\cite{dc1, dc2, d3, d4, dc5, Mozes05, dc6, Srivastava19}. For experimental realizations of QSD, see ~\cite{E1, E2, E3, E5, E6, E9, E10, E11, E4, E7, E12, E13, E8}.

Returning to the information processing aspects of QSD, an important extension arises when multiple receivers aim to decode the information. In such cases, each receiver performs a measurement to discriminate the states before passing them sequentially to the next receiver. This process, known as sequential state discrimination (SSD) ~\cite{Seq1, Seq2}, has been experimentally implemented in ~\cite{ESeq}. If the concerned states are bipartite, each receiver can be considered to be in a pair, where each member of the pair performs local measurements on the subsystem in possession and classically communicate with each other. This leads to SSD via local operations and classical communication (LOCC), which is experimentally feasible and desirable for quantum information processing. Unambiguous sequential state discrimination using local operations has been studied in ~\cite{US}. For MESD, ~\cite{ortho} demonstrated that any two pure orthogonal quantum states can be discriminated sequentially by an arbitrary number of pairs with unit success probability via LOCC. This was later extended to pure non-orthogonal states in ~\cite{nonortho}. {Additionally, ~\cite{ONon, OpSeq} provided conditions that optimal local measurements must satisfy to reach the success probability achievable through global measurements for SSD of any two bipartite quantum states}. The role of quantum correlations, such as discord, in SSD is studied in ~\cite{Disseq}. Further work on SSD can be found in ~\cite{Oseq1, Oseq2, Oseq3, Oseq4, Oseq5, Oseq6}.

In this work, {we address the problem of SSD of two entangled orthogonal pure states, prepared with equal prior probability.}  Previous work~\cite{ortho} has shown that such states can be perfectly discriminated by an arbitrary number of pairs of parties. However, the perfect discrimination process inevitably destroys the entanglement of the received states. Given that entanglement is a crucial resource for various quantum communication tasks, including teleportation~\cite{tl1,tl2}, dense coding~\cite{dc1,dc2,d3,d4,dc5,Mozes05,dc6,Srivastava19}, and that any two-qubit entangled state is always distillable~\cite{Dis}, preserving entanglement—even in small amounts—becomes an important objective. The idea of preserving resources while collecting useful information, sequentially, from given quantum systems has been an extensive area of recent research~\cite{Silva15,Mal16,Brown20,sriv21,WC,pandit22,Srivastava25,mondal23}. {Motivated by this, we go beyond conventional SSD and aim to develop a protocol that not only retrieves information through sequential discrimination, but also preserves a portion of the entanglement of the states involved. We refer to this approach as Sequential State Discrimination Sustaining Entanglement (SSDSE). 
Our protocol employs LOCC within the MESD paradigm. However, sustaining entanglement necessitates non-optimal measurements, which may reduce the success probability of discrimination. Despite this trade-off, our goal is to ensure that the quantum advantage of measurement remains superior to random guessing. }

  We first employ our protocol to the case of discriminating between two specific two-qubit orthogonal entangled pure states. We 
 show that the success probability surpasses the success probability of random guessing for an arbitrarily long sequence of state discrimination. Additionally, using logarithmic negativity~\cite{neg1,neg3,neg4,neg2} as a measure of entanglement, we confirm that these states retain a finite non-zero amount of entanglement throughout the process.
 In the latter part of our study, we extend our approach to encompass the discrimination of any two pure, entangled and orthogonal two-qubit states, each prepared with equal probability. We first demonstrate that an entanglement witness operator always exists to verify the presence of entanglement in individual states at the end of each state discrimination step. We show that witnessing entanglement by an arbitrary number of pairs using this operator requires each pair to make a specific choice of measurement parameters for state discrimination. Given such choices, we establish that the success probability of discrimination at each step always exceeds that of random guessing. Thus, our protocol achieves both objectives: discrimination better than random guessing and retention of entanglement. Our analysis holds for any number of participating parties, ensuring that an arbitrary number of parties can discriminate the states more effectively than random guessing without completely destroying their entanglement. 
 It is worth noting that the case of two general orthogonal entangled states differs from that of two specific orthogonal entangled states considered earlier. In the latter case, the success probability of discrimination at each step can be made arbitrarily close to one. This keeps non-zero entanglement in the post-measurement states for the next pair of sequential observers as long as the success probability remains less than one.   However, such a thing is not guaranteed in the general case.

 The rest of the paper is organized as follows. In Sec.~\ref{A}, we discuss the preliminaries, covering MESD in Sec.~\ref{1A}, MESD using LOCC in Sec. \ref{1B}. We discuss SSD and our protocol SSDSE in Sec.~\ref{sec3}. In Sec.~\ref{S4}, we consider an ensemble of two particular orthogonal entangled states and explicitly compute the discrimination probability at each step (Sec.~\ref{S4A}) and, using logarithmic negativity as an entanglement measure, show that the states remain entangled regardless of the number of participating parties (Sec.~\ref{S4B}).  Considering an ensemble of general two two-qubit orthogonal states, the SSDSE is elaborated in Sec.~\ref{S3}, which involves constructing a witness operator and detecting the entanglement of each state at every round (Sec.~\ref{S3A}) and analyzing the success probability of the protocol (Sec.~\ref{S3P}). lastly conclude in Sec.~\ref{con}

\section{Quantum state discrimination}
\label{A}
QSD~\cite{Re1,Re2,Re3,Re4,Re5,Re6} deals with the problem of distinguishing quantum states, randomly selected from a known ensemble with a particular probability distribution. 
We adopt MESD to quantify the extent of QSD. In MESD, the average probability of misidentification of states is minimized by choosing the optimal measurement settings. In other words, a measurement strategy is employed that gives a maximum average probability of successfully identifying the quantum states chosen from a given ensemble. 
\subsection{Minimum-error state discrimination protocol}
\label{1A}
Consider an ensemble of $n$ quantum states, denoted as $\eta_n := \{q_b, \zeta_b\}$, where $b \in \{1, 2, \dots, n\}$, $q_b> 0$ $\forall$ $b$, and $\sum_{b=1}^{n} q_b = 1$. In this ensemble, each state $\zeta_b$ is prepared with probability $q_b$ and is uniquely identified by the index $b$. Furthermore, we assume that the states act on a Hilbert space of finite dimension, $d$. A source selects a state, say $\zeta_b$, from the ensemble, $\eta_n$, and sends it to a receiver.
In a QSD game, it is assumed that both the source and the receiver have prior knowledge of this ensemble.  The receiver's goal is to identify the received state, relying solely on prior knowledge of the ensemble and applying preferred measurements on the received state. To achieve this, the receiver can implement a positive-operator-valued measurement (POVM), denoted as $\{T_i \geq 0\}$, where $i \in {1, 2, \dots, m}$ and $\sum_i T_i = \mathbb{I}_d$. Here and throughout the paper, we use the standard conventional notation, $\mathbb{I}_d$, to denote the identity operator that acts on a Hilbert space of dimension $d$.

The strategy is to guess the state to be $\zeta_b$, with probability $p(b|i)$, if the measurement outcome corresponding to $T_i$ clicks. This implies that the probability of a correct guess, when the received state is $\zeta_b$ and $i^\text{th}$ outcome is clicked, is $p(b|i)$. According to Born's rule, the probability of obtaining $T_i$, when the given state is $\zeta_b$, is $\Tr[\zeta_b T_i]$. Consequently, the average success probability of correctly discriminating the states in $\eta_n$ using the POVMs, $\{T_i\}$, is given by
\begin{equation*} P_{\text{suc}} =\sum_{i,b} q_b\Tr[\zeta_b T_i]p(b|i). \end{equation*}
To achieve the maximum success probability, $P_{\text{max}}$, one must optimize the average success probability, $P_{\text{suc}}$, over all possible measurement settings, $\{T_i\}$, and strategies, $\{p(b|i)\}$, i.e.,
\begin{eqnarray*}
    P_{\text{max}}&=&\max \limits_{\{T_{i}\}\in \mathcal{T},\{p(b|i)\}\in\mathcal{P}} P_{\text{suc}}, \hspace{1.9cm}\\
   &=&\max\limits_{\{T_{i}\}\in \mathcal{T},\{p(b|i)\}\in\mathcal{P}} \sum_{i,b} q_b\Tr[\zeta_b T_i]p(b|i),
\end{eqnarray*}
where $\mathcal{T}$ represents the set of all POVMs acting on $d$ dimensional Hilbert space and $\mathcal{P}$ is the set of all conditional probability distributions. 
\subsection{Minimum-error state discrimination using LOCC}
\label{1B}
When the ensemble consists of states involving two systems, the measurements to be performed on the selected state can act locally on individual systems, locally on individual systems along with classical communication between the parties holding the systems, or globally, depending on the available resources. The last case is the same as that discussed in the previous subsection, because the bipartite states can be considered as a single system when global measurements are allowed. Let us move to the second case, which is state discrimination using LOCC.

In this regard, consider an ensemble of bipartite states, $\eta_n:=\{t_b, \tau_b\}$, where $b \in \{1, 2, \ldots, n\}$,  $n \in \mathbb{N} (\geq 2)$, $t_b>0$ $\forall$ $b$, and $\sum_{b=1}^{n} t_b = 1$. Here, $t_b$ is the probability with which the bipartite quantum state, $\tau_b$, is prepared.
The states, $\{\tau_b\}$, which are to be discriminated, operates on the Hilbert space $\mathbb{C}^{d_1} \otimes \mathbb{C}^{d_2}$, where $d_1, d_2 \in \mathbb{N}$ ($d_1, d_2 \geq 2$) are the dimensions of the individual systems. The source picks a state, say $\tau_b$, from this ensemble, $\eta_n$, and, instead of sending $\tau_b$ to a single receiver, distributes one subsystem to, say, Alice and the other subsystem to, say, Bob. The receivers are aware of the ensemble, $\eta_n$, and collaboratively attempt to discriminate the states of it, i.e., to reveal the index $b$ of the received state. However, considering them to be situated in different locations, we restrict them to perform only local measurements and to communicate classically. 
Furthermore, the communication is restricted to be one-way, i.e., only one of them is permitted to communicate the measurement results to the other during the performance of the measurements. However, at the end of the action of all the measurements, both parties will discuss among themselves to make a guess about $b$, depending on the measurement outcomes. Without loss of generality let us consider that Alice makes the first measurement and informs Bob about her outcome and then Bob performs his measurement. We denote the set of local POVM elements of the measurement performed by Alice as $\{M_i \geq 0\}$, where $i \in \{1,2, \ldots,m_1\}$ and $
\sum_i M_i = \mathbb{I}_{d_1}$. Suppose the measurement outcome corresponding to $M_i$, labeled by $i$, is clicked. Alice then classically communicates her measurement outcome, $i$, to Bob. Depending on $i$, Bob performs a local POVM measurement, $\{N_{j|i} \geq 0\}$, where {$j \in \{1,2, \ldots, m_2^i\}$} and  
$\sum_{j} N_{j|i} = \mathbb{I}_{d_2}$ $\forall$ $i$. In this entire process, the probability that a particular pair of measurement outcomes, $\{i,j\}$, is clicked, when the received state is $\tau_b$, is  
$\operatorname{Tr}[\tau_b M_i \otimes N_{j|i}]$. Similar to the previous strategy, after the performance of the measurements, Alice and Bob collaboratively guess the state to be $\tau_b$ with probability $p(b|i,j)$, depending on the measurement outcomes $i$ and $j$. Hence the probability of a correct guess, given the measurement outcomes are $i$ and $j$, is $p(b|i,j)$.  
Hence, the average success probability corresponding to the one-way LOCC setting in this case can be expressed as
\begin{eqnarray*}
P_{\text{suc}}^{\text{ow}} &=&\sum_{i,j,b} t_b  \Tr[\tau_b M_i \otimes N_{j|i}] p(b|i,j),
\end{eqnarray*}
where the $\text{ow}$ in the superscript helps to keep in mind that the MESD process considered here involves only one-way LOCC. 
Therefore, the maximum average success probability in this case is simply  
\begin{eqnarray*}
P_{\text{max}}^{\text{ow}}
    = \max \limits_{\substack{\{M_i\otimes N_{j|i}\}\in \chi,\\ \{p(b|i,j)\}\in\mathcal P}}
    \sum_{i,j,b} t_b  \Tr[\tau_b M_i \otimes N_{j|i}]\\ p(b|i,j),
\end{eqnarray*}
where $\chi=\mathcal{M} \otimes \mathcal{N}$ and $\mathcal P$ is the set of all conditional probability distributions. Here, $\mathcal{M}$ and $\mathcal{N}$ denote the sets of all local POVMs acting on Hilbert spaces of dimensions $d_1$ and $d_2$, respectively.

In the next subsection, we discuss an example of MESD using LOCC, considering an ensemble of two orthogonal two-qubit pure states. 

\subsubsection{Example: LOCC discrimination of two pure orthogonal two-qubit states}
\label{2o}
 According to Ref.~\cite{ortho}, any two pure orthogonal two-qubit states, $\ket{\Phi_{1}}$ and $\ket{\Phi_{2}}$, up to local unitary, can be expressed as follows: 
\begin{align}
      \ket{\Phi_1} &= \sqrt{\mu_1} \ket{00}_{AB} + \sqrt{1-\mu_1} \ket{1\varsigma_1}_{AB},
   \nonumber \\ 
    \ket{\Phi_2} &= \sqrt{\mu_2} \ket{01}_{AB} + \sqrt{1-\mu_2} \ket{1\varsigma_1^\perp}_{AB}.
    \label{ortho1}
\end{align}  
Here $0 \leq \mu_1$, $\mu_2 \leq 1$, $\braket{0} {1}_{A/B} = 0$ and $\braket{\varsigma_1} {\varsigma_1^\perp}_B = 0$. 
Without loss of generality, we can consider
\begin{eqnarray}
\ket{\varsigma_{1}}_B &= \cos{\theta} \ket{0}_B + \sin{\theta} \ket{1}_B, \label{sabdhaner mar nei}\\
\ket{\varsigma_{1}^\perp}_B &= -\sin{\theta} \ket{0}_B + \cos{\theta} \ket{1}_B, \label{megh na chaite jol}  
\end{eqnarray}
where $\theta \in [0, \pi/2]$ is the angle made by the line joining the points $P(\ket{0}_B)$ and $P(\ket{1}_B)$ with the line joining the points $P(\ket{\varsigma}_B)$ and $P(\ket{\varsigma^\perp}_B)$. Here $P(\ket{\cdot})$ denotes the point on the Bloch sphere representing the qubit-state $\ket{\cdot}$. {The state is entangled if $\mu_{1,2}\in (0,1)$ and $\theta\in (0, \pi/2]$.} 
Consider an ensemble of two states, $\zeta_2=\{p_{b},\rho_b\}$, where $\rho_b=\ketbra{{{\Phi}}_b}{{\Phi}_b}$, with $b \in \{1,2\}$. 
Ref.~\cite{ortho} has shown that the maximum average success probability of discriminating states from this known ensemble is always unity, even when the parties, Alice and Bob, are restricted to perform only one-way LOCC operations. Let us describe this optimal protocol in more detail. Suppose the set of optimal LOCC measurements are $\{A_i\otimes B_{i|j}\}_{i,j=0}^{1}$ and the optimal probability distribution, $\{p(b|i,j)\}$, is $\{p_\text{opt}(b|i,j)\}$.
 For the concerned ensemble, the elements of the optimal LOCC measurement are 
 \begin{eqnarray}
  A_0&=&\ketbra{0}{0}_A\text{, }A_1=\ketbra{1}{1}_A\text{, }\nonumber\\B_{0|0}&=&\ketbra{0}{0}_B\text{, }B_{1|0}=\ketbra{1}{1}_B\text{, }\nonumber\\
  B_{0|1}&=&\ketbra{\varsigma_1}{\varsigma_1}_B\text{, }B_{1|1}=\ketbra{\varsigma_1^\perp}{\varsigma_1^\perp}_B.\label{myeq1}   
 \end{eqnarray}
 Clearly, after the measurement, if detection occurs at $A_0\otimes B_{0|0}$ or $A_1\otimes B_{0|1}$, the state can be correctly identified as $\rho_1$. Similarly, if detection occurs at $A_0\otimes B_{1|0}$ or $A_1\otimes B_{1|1}$, the state is surely $\rho_2$. Hence $p_\text{opt}(1|0,0)=p_\text{opt}(1|1,0)=p_\text{opt}(2|0,1)=p_\text{opt}(2|1,1)=1$ and $p_\text{opt}(b|i,j)=0$ otherwise. 
 Therefore the average success probability can easily be checked to be 1.


 
\section{Sequential one-way LOCC state discrimination while  sustaining entanglement} 
\label{sec3}
SSD is a quantum communicational task in which multiple receivers sequentially attempt to retrieve information from a state selected from a given ensemble by performing measurements on the state. In this section, we briefly explain the SSD protocol considering ensembles of bipartite states. The concept can be easily generalized to multipartite states.


Consider a sequence of $K$ pairs of receivers, Alice$_k$ and Bob$_k$, where $k\in\{1,2,\ldots,K\}$. Each Alice$_k$ is specially separated from the corresponding Bob$_k$. Similar to the scenario described in the previous subsection, a source randomly selects a bipartite state, $\tau_b$, from a known ensemble, $\eta_n:=\{t_b,\tau_b\}_{b=1}^n$, and sends one subsystem of $\tau_b$ to Alice$_1$ and the other to Bob$_1$. The SSD protocol with one-way LOCC for such an ensemble can then be described as follows:

\begin{itemize}
\item Alice$_1$ performs a local POVM measurement on her part of the received state, $\tau_b$, and classically communicates the measurement outcome to Bob$_1$, based on which he performs a local POVM measurement on his part of $\tau_b$. After finishing both of the measurements, they together try to discriminate the state, $\tau_b$. Let the final post-selected state, after the application of the optimal measurement, be $\tau_b^1$.
\item Afterwards, Alice$_1$ sends her part of $\tau_b^1$ to Alice$_2$, and Bob$_1$ sends his part of $\tau_b^1$ to Bob$_2$. {Alice$_1$ and Bob$_1$ do not communicate their measurement outcomes with the next sequential pair Alice$_2$ and Bob$_2$.} Alice$_2$ and Bob$_2$ then apply a one-way LOCC measurement scheme on $\tau_b^1$ to distinguish the states of the initial ensemble, $\eta_n$. Let the final state after the application of the measurements be $\tau_b^2$. One may need to keep in mind that $\tau_b^k$ will depend on the measurement outcomes of all Alice$_{\tilde{k}}$ and Bob$_{\tilde{k}}$ with $\tilde{k}\leq k$.
\item This process continues up to $K$ rounds.
\end{itemize}

\begin{figure*}
\vspace{-3cm}
\hspace{-2cm}
\includegraphics[scale=0.37]{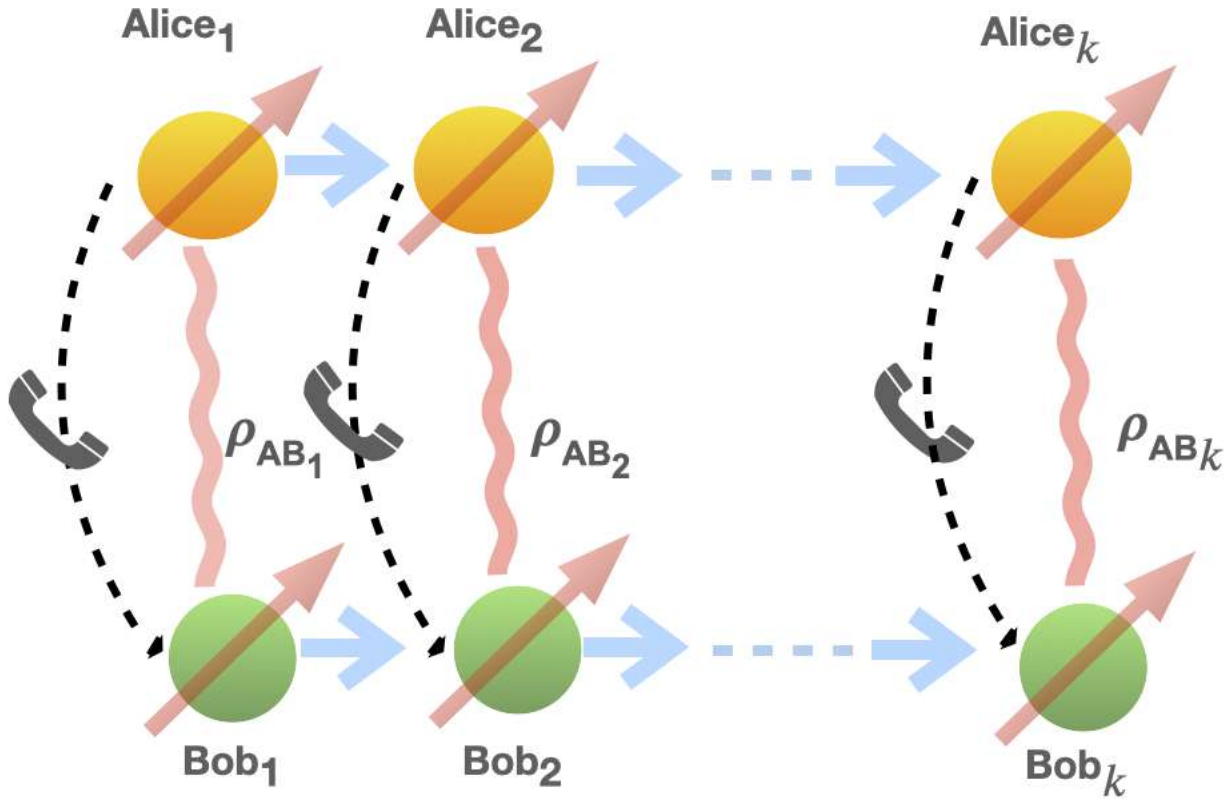}\hspace{2cm}
\includegraphics[scale=0.37]{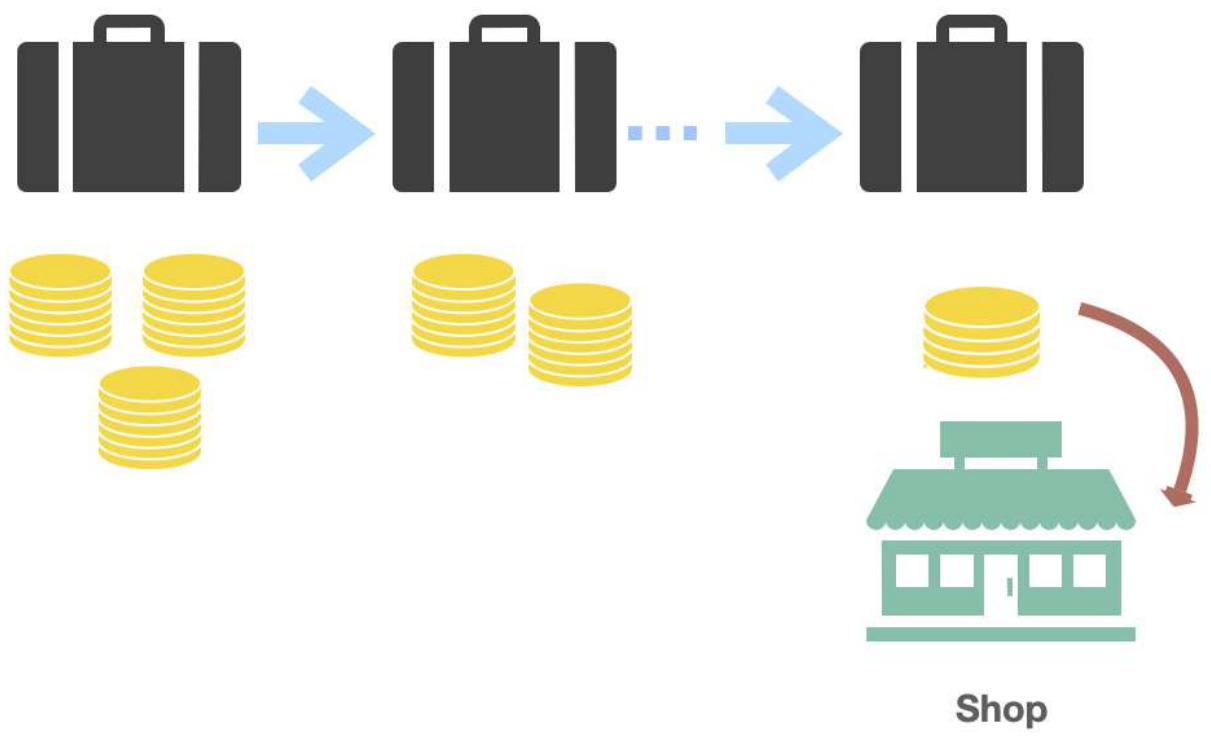}
\hspace{-2cm}
\vspace{-3cm}
\caption{
\textbf{
The SSD process using one-way LOCC protocol and a real-world interpretation of the SSDSE protocol.} The left part of the figure illustrates the SSD protocol schematically. As shown, the first pair—Alice$_1$ and Bob$_1$—initially share a bipartite quantum state, with Alice$_1$ holding one subsystem (colored orange) and Bob$_1$ holding the other (colored green). To discriminate the state, Alice$_1$ performs a measurement on her part and communicates the outcome to Bob$_1$ via a classical channel, depicted in the figure as a dotted arrow with a phone icon, indicating one-way classical communication. Based on the outcome received from Alice$_1$, Bob$_1$ performs a measurement on his subsystem. After both Alice$_1$ and Bob$_1$ have performed the measurements, they pass their respective subsystems on to the next pair, Alice$_2$ and Bob$_2$, with Alice$_2$ now holding the orange-colored subsystem and Bob$_2$ holding the green-colored subsystem. This process repeats sequentially, with each subsequent pair continuing the discrimination task, until the $k^\text{th}$ pair successfully identifies the state. {The right part of the figure illustrates the central concept of the SSDSE protocol through a real-life analogy. In this depiction, the black suitcase represents a quantum state that is sequentially passed among a series of receivers. The yellow coins symbolize entanglement. The suitcase contains important documents. To access its contents, each party must spend some coins to purchase a key. The first receiver on the left uses a portion of the coins to unlock the suitcase and read the documents—analogous to the state discrimination task performed by the first pair. After reading, they secure the documents back inside, relock the suitcase, and pass it to the next receiver. Each subsequent receiver repeats this process: spending a portion of the remaining coins to unlock the suitcase, accessing the documents, and then passing it along. Crucially, each ensures that enough coins remain for the next party to obtain a key. Eventually, the 
final receiver, who does not intend to read the documents, utilizes the residual coins to purchase essentials from a shop. This mirrors the use of remaining entanglement by the last party in a sequential state discrimination protocol to perform another quantum information task, such as quantum teleportation.}}
\label{f1}
\end{figure*}

A pictorial representation of the SSD protocol using one-way LOCC is provided in the left panel of Fig.~\ref{f1}. 

For the ensemble, $\zeta_2$, discussed in the previous section, it is evident that using the same set of POVM measurements, $\{A_i\otimes B_{i|j}\}_{i,j=0}^{1}$ (defined in \eqref{myeq1}), the SSD process can be continued for an arbitrary number of rounds, $K$, with each pair of Alice$_k$ and Bob$_k$ being able to locally discriminate the states of $\zeta_2$ with certainty.

However, it is important to note that after the initial measurements applied by Alice$_1$ and Bob$_1$, the output state, $\rho_b^1$, no longer remains entangled, even if the initial state, $\rho_b$, is entangled. Since LOCC can not increase entanglement, afterwards, the bipartite state shared between Alice$_l$ and Bob$_l$ remains unentangled throughout the sequential protocol, for all $1\leq k\leq K$. 

Entanglement is a fundamental resource in quantum communication, enabling advantages in various tasks that are unattainable with classical resources~\cite{tl1,tl2,dc1,dc2,d3,d4,dc5,Mozes05,dc6,Srivastava19}. However, standard LOCC QSD processes often lead to its complete dissolution. Thus, in this work,  recognizing the crucial role of entanglement as a resource in various quantum tasks, we want to discriminate bipartite states sequentially, avoiding the complete destruction of entanglement. 
Therefore, we replace the sharp measurements performed by each party of the $k^{\text{th}}$ pair with the corresponding unsharp measurements. We define an unsharp measurement, $\{Q_j\}$, corresponding to a rank-1 projective measurement, $\{P_j\}$ such that $\sum_j P_j = \mathbb{I}_d$, as
\begin{equation}
    Q_j = \lambda P_j + (1 - \lambda)\frac{\mathbb{I}_d}{d},
    \label{unsharp}
\end{equation}
where $\lambda \in (0,1)$ is the sharpness parameter that determines the degree of white noise mixed with $P_j$ and $d$ is the dimension of the system on which measurements are being applied. 

 We refer to our protocol as sequential state discrimination sustaining entanglement. {The right panel of Fig.~\ref{f1} schematically illustrates a physical analogy to the scenario under consideration. The figure depicts a locked black suitcase filled with important documents, which is passed sequentially to multiple receivers. This suitcase symbolizes the quantum state, while the yellow coins represent entanglement. Each receiver is permitted to utilize a portion of the coins (symbolizing entanglement) to acquire a key to unlock the suitcase and read the documents—analogous to the state discrimination task. However, they must ensure that some of the coins remain unused; after reading the document, they place it back into the suitcase, relock it, and pass it, along with the remaining coins, to the next receiver. This process continues until the final receiver, who is not interested in reading the document, uses the remaining coin to purchase essentials (represented by a shop in the figure).} 
The objective of SSDSE closely mirrors this analogy. 


In the next section, we employ the SSDSE protocol in a specific case involving an ensemble of two well-defined entangled pure states, prepared with equal prior probability. By applying the SSDSE protocol to this ensemble, we compute the exact entanglement of the post-measurement states at each round, quantified via logarithmic negativity, and confirm that it remains nonzero throughout an arbitrary number of rounds. Moreover, we compute the corresponding success probabilities, which consistently exceed the success probability of a random guess, $\frac{1}{2}$. 

{Later, we apply the SSDSE protocol to a more generic ensemble, and show that both the objectives of sustaining entanglement and discrimination better than random guessing always hold true.}

\section{Reaching arbitrarily high success probability while sustaining entanglement}
\label{S4}

In this section, we consider the SSDSE protocol applied to an ensemble $\chi_2 = \{p_b, \bar{\rho}_b\}$, where $i =$1, 2 and $p_1 = p_2 = \frac{1}{2}$.   The states $\bar{\rho}_b$ are pure entangled states of the form $\bar{\rho}_b = \ketbra{\kappa_b}{\kappa_b}$, with  
\begin{eqnarray*}
   \ket{\kappa_1}&= \sqrt{\gamma_1}\ket{01} + \sqrt{1 - \gamma_1}\ket{10},\\
   \ket{\kappa_2}&= \sqrt{\gamma_2}\ket{00} + \sqrt{1 - \gamma_2}\ket{11},
\end{eqnarray*}
where $\gamma_{1,2} \in (0,1)$. These states are specific examples of pair of two-qubit pure orthogonal states. 


In the following subsections, we state and prove a theorem regarding the average success probability of discriminating $\ket{\kappa_1}$ and $\ket{\kappa_1}$. Moreover, we determine the entanglement of the output states after the application of measurements at each round.

\subsection{Probability of discrimination in each round}
\label{S4A}

{Each pair involved in the protocol is aware of the states present in the ensemble $\chi_2$. When a pair receives a state for identification, they could, in principle, make a completely random guess---e.g., by tossing a fair coin. If a head appears, they guess the state to be $\rho_1$, and if it is a tail, they guess $\rho_2$. This approach is entirely measurement-independent. Since each state occurs with equal probability $\frac{1}{2}$, and the guessing strategy is random, the probability of correctly identifying the state is:
$P_R=2 \times \frac{1}{2} \times \frac{1}{2} = \frac{1}{2}$, where the subscript $R$ denotes that it represents the probability of success through a random guess.}

{The goal of the parties is to outperform $P_R$ by employing measurements to identify the states, thereby gaining an advantage. However, they also want to sustain the entanglement of the states during the process. Hence, they implement a measurement scheme that is not optimal but yields a success probability greater than $P_R=\frac{1}{2}$. Below we present a theorem that ensures that using SSDSE protocol every pair at $k^{\text{th}}$ round can discriminate the states in $\chi_2$ with success probability greater than $P_R$.}

The optimal measurement for the perfect discrimination of the states of $\chi_2$ is given by

\begin{align*}
    \bar{O}_1 &= \ketbra{0}{0}_A \otimes \ketbra{1}{1}_B, \\
    \bar{O}_2 &= \ketbra{0}{0}_A \otimes \ketbra{0}{0}_B, \\
    \bar{O}_3 &= \ketbra{1}{1}_A \otimes \ketbra{0}{0}_B, \\
    \bar{O}_4 &= \ketbra{1}{1}_A \otimes \ketbra{1}{1}_B.
\end{align*}
If the detection occurs at $\bar{O}_1$ or $\bar{O}_3$,  the state is identified as $\rho_1$, whereas a detection at $\bar{O}_2$ or $\bar{O}_4$ corresponds to identifying the state as $\rho_2$. 

This set of measurements will completely destroy the entanglement. Hence, we introduce a key constraint on the allowed measurements: each pair must ensure that their discrimination process does not completely eliminate the entanglement of the states they are distinguishing. As a result, by the end of the sequential state discrimination process, the final pair retains a finite amount of entanglement, which can be used for various quantum information processing tasks.
To maintain this constraint, instead of measuring with $\{ \bar{O}_1\}_{j=1}^{4}$ the receivers measure with unsharp measurements.



To sequentially distinguish the states of $\chi_2$ while sustaining its entanglement, we implement the SSDSE protocol by replacing the sharp projective measurements $\bar{O}_j$ with the corresponding unsharp measurements $\bar{O}_j^k$ in each round $k$ as given in Eq.~\eqref{unsharp}. The explicit forms of $\bar{O}_j^k$ are presented below:

\begin{equation}
\begin{aligned}
    \bar{O}_1^k &= \dfrac{1}{4}\Big[(1+\bar{\lambda}_k)^2\ketbra{01}{01} + (1-\bar{\lambda}_k)^2\ketbra{10}{10} \\ 
    &\quad + (1-\bar{\lambda}_k^2)(\ketbra{00}{00} + \ketbra{11}{11})\Big], \\
      \bar{O}_2^k &= \dfrac{1}{4}\Big[(1+\bar{\lambda}_k)^2\ketbra{00}{00} + (1-\bar{\lambda}_k)^2\ketbra{11}{11} \\
    &\quad + (1-\bar{\lambda}_k^2)(\ketbra{01}{01} + \ketbra{10}{10})\Big],\\
    \bar{O}_3^k &= \dfrac{1}{4}\Big[(1+\bar{\lambda}_k)^2\ketbra{10}{10} + (1-\bar{\lambda}_k)^2\ketbra{01}{01} \\
    &\quad + (1-\bar{\lambda}_k^2)(\ketbra{00}{00} + \ketbra{11}{11})\Big].\\
     \bar{O}^k_4 &= \dfrac{1}{4}\Big[(1+\bar{\lambda}_k)^2\ketbra{11}{11} + (1-\bar{\lambda}_k)^2\ketbra{00}{00} \\
    &\quad + (1-\bar{\lambda}_k^2)(\ketbra{10}{10} + \ketbra{01}{01})\Big], \\
    \label{Mes}
\end{aligned}
 \end{equation}
Here, $\bar{\lambda}_k$ denotes the sharpness parameter of the measurements used in the $k^\text{th}$ round. One can easily check that this POVM can be implemented through LOCC. Following the same strategy as in the optimal case, if either $\bar{O}_1^{k}$ or $\bar{O}_3^{k}$ clicks, the state is identified as $\bar{\rho}_1$, and if $\bar{O}_2^{k}$ or $\bar{O}_4^{k}$ clicks, the state is identified as $\bar{\rho}_2$. Using this measurement setting, we sequentially discriminate the states in $\chi_2$. Below, we present a theorem that ensures the success probability in each round always exceeds the random guessing probability  $P_R = \frac{1}{2}$.

{\begin{theorem}
\label{th1}
    The success probability of discriminating the states in $\chi_2$ at an arbitrary round $k$, using the SSDSE protocol, is always greater than the random guessing probability $P_R$, irrespective of the number of pairs participating in the discrimination protocol, and is given by
\begin{equation*}
    \bar{P}_{\mathrm{suc}}^k = \frac{1}{2} + \frac{1}{2}\bar{\lambda}_k^2.
\end{equation*}
\end{theorem}}

\begin{proof}


{The average success probability for the first round, $k = 1$, can be computed as}
\begin{equation}
    \bar{P}_\text{suc}^1 = \frac{1}{2} \left( \Tr\left[(\bar{O}_1^1 + \bar{O}_3^1)\bar{\rho}_1\right] + \Tr\left[(\bar{O}_2^1 + \bar{O}_4^1)\bar{\rho}_2\right] \right).
\end{equation}
{Substituting the expressions for $\bar{\rho}_1$, $\bar{\rho}_2$, and $\bar{O}_j^1$ for $j = 1, 2, 3, 4$, we obtain}
\begin{equation}
    \bar{P}_\text{suc}^1 = \frac{1}{2} + \frac{1}{2} \lambda_1^2.
\end{equation}

To compute the average success probability for the subsequent round, $k>1$, we first determine the post-measurement state at an arbitrary $(k-1)^\text{th}$ round. For this, we recall that if the state at the end of $(k-2)^\text{th}$ round is $\bar{\rho}_{b}^{k-2}$ then the post-measurement state after $(k-1)^\text{th}$ round by using the von Neumann-L\"uder's rule \cite{Busch}
\begin{equation}
    \bar{\rho}_{b}^{k-1} = \sum_j \sqrt{\bar{O}_j^{k-1}} \,\bar{\rho}_{b}^{k-2} \, \sqrt{\bar{O}_j^{k-1}},
    \label{sta}
\end{equation}
where $b =$ 1, 2 corresponds to the first and second states of the ensemble $\chi_2$.

By recursively using Eqs.~\eqref{Mes} and~\eqref{sta} we obtain the post-measurement states at the end of $k^\text{th}$ round in terms of the initial states $\bar{\rho}_{1}$ and $\bar{\rho}_{2}$ as:

{\begin{equation}
\begin{aligned}
    \bar{\rho}_{1}^{k-1} &= \bar{\rho}_1 - S_{k-1} \left(\ketbra{01}{10} + \ketbra{10}{01}\right) \vartheta_1, \\
    \bar{\rho}_{2}^{k-1} &= \bar{\rho}_2 - S_{k-1} \left(\ketbra{00}{11} + \ketbra{11}{00}\right) \vartheta_2,
    \label{rstnew}
\end{aligned}
\end{equation}
where
\begin{eqnarray}
S_1&=& \bar{\lambda}_1^2, \nonumber \\ S_l&=&\bar{\lambda}_l^2 + S_{l-1}(1 - \bar{\lambda}_l^2)~~ \forall l>1, 
    \label{recur}
\end{eqnarray}
and $\vartheta_1 = \sqrt{\gamma_1(1 - \gamma_1)}$, $\vartheta_2 = \sqrt{\gamma_2(1 - \gamma_2)}$.  
}

Once we have obtained the post-measurement states in an arbitrary round $k-1$, we can compute the average success probability of discriminating the states at the $k^\text{th}$ round as:
\begin{equation}
    \bar{P}_\text{{suc}}^k = \frac{1}{2}\left( \Tr\left[(\bar{O}_1^k + \bar{O}_3^k)\bar{\rho}_1^{k-1}\right] + \Tr\left[(\bar{O}_2^k + \bar{O}_4^k)\bar{\rho}_2^{k-1}\right] \right).\label{myeq5}
\end{equation}
Note that $\bar{P}_\text{suc}^k$ depends on the post-measurement state, $\bar{\rho}_b^{k-1}$, of the $(k-1)^\text{th}$ round, because the receivers of $k^\text{th}$ round receive $\bar{\rho}_b^{k-1}$ and perform a LOCC measurement on it to discriminate between the two states of the ensemble $\chi_2$.

{Thus, by substituting the expressions of $\bar{O}_j^k$, $\bar{\rho}_1^{k-1}$, and $\bar{\rho}_2^{k-1}$, given in Eqs.~\eqref{Mes} and ~\eqref{rstnew}, in Eq. \eqref{myeq5}, we obtain the average success probability of discrimination at the $k^\text{th}$ round as:
\begin{equation*}
    \bar{P}_\text{{suc}}^k = \frac{1}{2} + \frac{1}{2}\bar{\lambda}_k^2.
\end{equation*}}
This completes the proof of Theorem~\ref{th1}.
\end{proof}
Note that $\bar{P}_\text{{suc}}^k = 1$ if the pair at the $k^\text{th}$ site chooses $\bar{\lambda}_k = 1$. This occurs because the states $\bar{\rho}_1^{k-1}$ and $\bar{\rho}_2^{k-1}$ always lie in orthogonal subspaces, and can therefore be perfectly discriminated using LOCC.

Hence from Theorem.~\ref{th1} we infer that $\bar{P}_\text{{suc}}^k$ is always greater than $P_R$ for any non-zero $\bar{\lambda}_k$, regardless of the value of $k$, as long as $k \in \mathbb{N}$. It is also worth noting that the discrimination probability at each step, $\bar{P}_\text{{suc}}^k$, depends only on the sharpness parameter of the corresponding measurement and is independent of the sharpness parameters from preceding rounds. 

Note that if the $k^\text{th}$ pair chooses $\bar{\lambda}_k = 0$, the protocol becomes equivalent to random guess in terms of success probabilities. On the other extreme, setting $\bar{\lambda}_k = 1$ results in perfect state discrimination. However, this would completely destroy the entanglement of the states, as we will demonstrate in the next subsection.  

Once the criterion of enhanced discrimination probability is satisfied, the next step is to evaluate the entanglement of the states at each round for $\bar{\lambda}_k \in (0, 1)$, to ensure that the entire process adheres to the SSDSE protocol. This analysis is presented in a form of a theorem in the following subsection.

\subsection{Residual entanglement after each round}
\label{S4B}
In this subsection, we compute the entanglement of the individual states, $\bar{\rho}_b^k$, at each step, $k$, of the SSDSE protocol. {We present the following theorem that suggests that the individual states of $\chi_2$ remain entangled at each measurement step.
\begin{theorem}
   The entanglement of the state, $\bar{\rho}^k_b$, after performing measurements in the $k$th round, quantified with logarithmic negativity, is
   \begin{equation}
E_b^k = \log_2 \left[ 1 + 2 \vartheta_b (1 - S_k) \right],
\end{equation}
 which is non-zero for $0<\gamma_b<1$ and  $0\leq\bar{\lambda}_l<1$ for $l=1,2,\ldots,k$. 
\end{theorem}}

\begin{proof}
We use logarithmic negativity~\cite{neg1,neg2,neg3,neg4} as the entanglement measure. For a bipartite state $\rho_{AB}$ in $\mathbb{C}^2\otimes \mathbb{C}^2$, the logarithmic negativity is given by:
\begin{equation*}
    E_n = \log_2(2\mathcal{N} + 1),
\end{equation*}
where $\mathcal{N} = \sum_{e_p < 0} |e_p|$ is the absolute sum of the negative eigenvalues, $e_p$, of the partially transposed density matrix $\rho_{AB}^{\Gamma_{A/B}}$. Here, $\Gamma_{A/B}$ denotes the partial transpose with respect to subsystem $A$ or $B$.

Using the form of the post-measurement state $\bar{\rho}_b^k$, as given in Eq.~\eqref{rstnew}, it is easy to check that the partial transposition of these states always renders a negative eigenvalue. The logarithmic negativity, $E_1^k$ and $E_2^k$, of the states, $\bar{\rho}_1^k$ and $\bar{\rho}_2^k$, is given by
\begin{equation*}
    E_1^k = \log_2 \left( 2 \mathcal{N}_1^k + 1 \right), \quad E_2^k = \log_2 \left( 2 \mathcal{N}_2^k + 1 \right),
\end{equation*}
where $\mathcal{N}_1^k$ and $\mathcal{N}_2^k$ represent the absolute sum of the negative eigenvalues of the partially transposition of  $\bar{\rho}_1^k$  and $\bar{\rho}_2^k$, respectively. Thus we get,
\begin{eqnarray*}
    E_1^k &= \log_2 \left[ 1 + 2 \vartheta_1 (1 - S_k) \right], \\
    E_2^k &= \log_2 \left[ 1 + 2 \vartheta_2 (1 - S_k) \right],
\end{eqnarray*}
where $S_k$ has the form as given in Eq.~\eqref{recur}. 
{Notice that $0<2\vartheta_b\leq 1$ for $0<\gamma_b<1$. Also,  $0\leq S_1 \leq S_2 \leq\ldots \leq S_{k-1} \leq S_k < 1$ for $\bar{\lambda}_{1,\ldots,k} \in (0,1)$. }
It is thus clear that the post-measurement states of any round $k$, regardless of the value of $k$, as long as $k \in \mathbb{N}$, always possess a finite amount of entanglement, i.e., $E_1^k, E_2^k > 0$, for $\bar{\lambda}_{1,\ldots,k} \in (0,1)$. Notably, if at any step the measurement is made sharp by choosing $\bar{\lambda}_k = 1$, then $S_k=1$, and thus the states no longer remain entangled, as expected. {Instead, if a $\bar{\lambda}_k \approx 1$, then it is still possible to get success probability close to one while keeping some amount of entanglement in the post measurement states.}   On the other hand, if $\bar{\lambda}_{k} = 0$ for all $k$, then the entanglement of the states remains unchanged from its initial value. However, in this case, the parties do not gain any advantage in success probability over the case of random guess, as discussed in the previous section. This completes the proof.
\end{proof}

{Thus, our analysis of the application of the SSDSE protocol to the specific case of discriminating two orthogonal pure two-qubit entangled states suggests that both objectives—namely, achieving a discrimination probability greater than that of random guessing and sustaining entanglement—are well satisfied in this case. Having established this, we now move on to extend our protocol to more generic ensembles of two orthogonal pure entangled two-qubit states, as discussed in the section below.
}


\section{The general case: SSDSE for two arbitrary orthogonal pure two-qubit states}
\label{S3}


{Let us now focus on a more generic ensemble, $\zeta_2$, of two general orthogonal pure two-qubit states, as discussed in Sec.~\ref{2o}. We fix $p_b = \frac{1}{2}$ for $b = 1$ and $2$, and refer to the resulting ensemble with fixed probabilities as $\eta_2$.}
 Since we are interested in sustaining entanglement, we will consider $\mu_b\in (0,1)$ and $\theta\in (0, \pi/2]$ so that the initial state, $\rho_b$, is entangled for both $b=1$ and 2. The task is to sequentially discriminate the states of $\eta_2$ such that each time the average success probability of discrimination is higher than the probability of discrimination in a random guess and the states after the measurements are still entangled.
The optimal measurement for the perfect discrimination of the states of $\eta_2$ is $\{A_i\otimes B_{i|j}\}_{i,j=0}^{1}$, as discussed in Sec.~\ref{2o}. This set of measurements will completely destroy the entanglement. {Hence we replace the optimal measurement with the corresponding unsharp measurements as defined in Sec. \ref{unsharp}}.


 To sequentially distinguish the states of $\eta_2$, sustaining its entanglement, the unsharp measurements that we use in the $k^\text{th}$ round are given by
\begin{widetext}
  \begin{align}
&A_0^{k} \coloneqq (1-\lambda_k)\frac{\mathbb{I}_2}{2}+\lambda_k A_0\text{, }A_1^{k} \coloneqq (1-\lambda_k)\frac{\mathbb{I}_2}{2}+\lambda_k A_1\text{, }
B_{0|0}^{k} \coloneqq (1-\lambda_k)\frac{\mathbb{I}_2}{2}+\lambda_k B_{0|0},\label{myeq4} \\ &B_{1|0}^{k} \coloneqq (1-\lambda_k)\frac{\mathbb{I}_2}{2}+\lambda_k B_{1|0}\text{, }
B_{0|1}^{k} \coloneqq (1-\lambda_k)\frac{\mathbb{I}_2}{2}+\lambda_k B_{0|1}\text{, }B_{1|1}^{k} \coloneqq (1-\lambda_k)\frac{\mathbb{I}_2}{2}+\lambda_kB_{1|1}\text{, }
\label{lo}
\end{align}  
\end{widetext}
Here, we consider that both parties in each pair of receivers at the $k^\text{th}$ round perform measurements characterized by the same sharpness parameter, $\lambda_k$.

Recall that we denote the two pure, entangled, and orthogonal states in the ensemble $\eta_2$ by $\rho_b$, with $b = 1$, $2$. Similarly, we denote the post-measurement state after the $k^\text{th}$ pair has performed their measurement as $\rho_b^k$, where $b = 1$, $2$ corresponds to the two individual states in the known ensemble $\eta_2$.

We also denote the measurement performed by the $k^\text{th}$ pair on the post-measurement state of the $(k-1)^\text{th}$ pair by $\{O_j^k\}$, where $j = 1,2,3,4$. Note that, according to this notation, $\rho_b^0$ represents the initial state selected from the ensemble $\eta_2$, which we will simply denote as $\rho_b$, with $b = 1,2$, throughout the rest of the paper. The elements of $\{O_j^k\}$ are given by:
\begin{eqnarray}
\label{jalsaghor} 
&O_{1}^k \coloneqq A_{0}^k\otimes B_{0|0}^{k}, \hspace{5 mm}
O_{2}^k \coloneqq A_{0}^k\otimes B_{1|0}^{k},\nonumber \\
&O_{3}^k \coloneqq A_{1}^k\otimes B_{0|1}^{k}, \hspace{5 mm}
O_{4}^k \coloneqq A_{1}^k \otimes B_{1|1}^{k} , 
\end{eqnarray}
As discussed before, 
the elements of $O_j^k$ forms an one-way LOCC measurement setting for all $k \in \mathbb{N}$.


{Using the measurement settings $\{O_j^k\}$ in the $k^\text{th}$ round, we analyse if the post-measurement states remain entangled.} To this end, we recall that the post-measurement state at the $k^\text{th}$ round can always be expressed in terms of the post-measurement state at the $(k-1)^\text{th}$ round as: 
\begin{equation}
\rho^{k}_{b} = \sum_{j=1}^4 \sqrt{O^k_j} \, \rho_b^{k-1} \, \sqrt{O^k_j},
\label{st}
\end{equation}
where $b = 1$, $2$ and $k \in \mathbb{N}$. The explicit forms of the operators $\sqrt{O^k_j}$, for $j = 1$, 2, 3, 4, are provided in Appendix~\ref{kya bolti h public}. By substituting these forms of $\sqrt{O^k_j}$ into Eq.~\eqref{st}, the post-measurement states $\rho_b^k$ at the $k^\text{th}$ round can be explicitly written in terms of the corresponding post-measurement states $\rho_b^{k-1}$ from the $(k-1)^\text{th}$ round. The detailed form of this recursive relation is also presented in Appendix~\ref{kya bolti h public}.


In the following subsection, we demonstrate how an appropriate choice of the unsharp parameter $\lambda_k$ can ensure that the post-measurement state, $\rho_b^k$, remains entangled at the end of round $k$. 

\subsection{Witnessing entanglement in each round}
\label{S3A}

Every pair at the $k^\text{th}$ round must ensure that their measurement does not completely destroy the entanglement of the received states $\rho^{k-1}_b$. To achieve this, each pair must carefully choose an appropriate sharpness parameter. But what should this choice be?

We address this question by proposing a witness operator that, through verifying the entanglement of the received state, guides how the sharpness parameter should be chosen. Specifically, we demonstrate that with a particular choice of the sharpness parameter at each round, it is possible to ensure that an arbitrarily large number of pairs can continue to discriminate the states without completely destroying their entanglement.


In Refs.~\cite{W1,W2,WC}, the authors showed that any operator on $\mathbb{C}^2 \otimes \mathbb{C}^2$ of the form
\begin{equation}
    \mathbb{W} = \frac{1}{4} \left[\mathbb{I}_2 \otimes \mathbb{I}_2 + \sum_{q=1}^3 g_q \, \sigma_q \otimes \sigma_q\right],
    \label{wit}
\end{equation}
can play a role of an entanglement witness for all $0 \leq \lvert g_q \rvert \leq 1$, since for every separable state, $\Sigma_S \in \mathbb{S}_{\text{sep}}$, it holds that $\Tr[\mathbb{W} \Sigma_S] \geq 0$. Here, $\mathbb{S}_{\text{sep}}$ denotes the set of all separable states on $\mathbb{C}^2 \otimes \mathbb{C}^2$ and $\sigma_q$ are the Pauli matrices.

Adopting this framework, we show existence of  witness operators to detect the entanglement of the post-measurement states $\rho^k_b$. The witness operator used at the $k^\text{th}$ round is defined as:
\begin{equation}
    \mathbb{W}^k_b = \frac{1}{4} \left[\mathbb{I}_2 \otimes \mathbb{I}_2 + \sigma_3 \otimes \sigma_3 - g_2^k \, \sigma_2 \otimes \sigma_2\right].
\end{equation}
{This operator has the same form as given in Eq.~\eqref{wit}, with $g_1 = 0$, $g_3 = 1$, and $g_2 = g_2^k$, where $0 \leq g_2^k \leq 1$. The superscript $k$ in $g_2^k$ indicates that the value of $g_2^k$ depends on the states being witnessed for entanglement in the $k^\text{th}$ round.} 
Such a witness operator $\mathbb{W}^k_b$ can certify the post-measurement state, $\rho^k_b$, as entangled if the condition $\Tr[\mathbb{W}^k_b \rho^k_b] < 0$ is satisfied.
This leads to the following inequality:
\begin{equation}
g_2^k > \frac{1 + \Tr[\sigma_3 \otimes \sigma_3 \, \rho^k_b]}{\Tr[\sigma_2 \otimes \sigma_2 \, \rho^k_b]},  \label{myeq2}  
\end{equation}
which must hold for all $k \in \{1,2,\ldots, K\}$ and $b \in \{1,2\}$ to confirm the retention of entanglement at the end of the last round.

{
It is possible to detect entanglement in the states $\rho^k_b$ if the sequence, $g^k_2$, is defined as
\small
\begin{equation}
\begin{aligned}
g_2^k 
&:= \left(1 + \epsilon_k\right) \max \left\{
\dfrac{1 + \Tr[\sigma_3 \otimes \sigma_3 \, \rho_1^k]}{\Tr[\sigma_2 \otimes \sigma_2 \, \rho_1^k]},
\dfrac{1 + \Tr[\sigma_3 \otimes \sigma_3 \, \rho_2^k]}{\Tr[\sigma_2 \otimes \sigma_2 \, \rho_2^k]}
\right\},
\label{sq}
\end{aligned}
\end{equation}
\normalsize
for $k=1,2,\ldots,K.$
Here, $\epsilon_k$ is a positive real number that depends on $k$ and is introduced to ensure that inequality~\eqref{myeq2} is satisfied, thereby guaranteeing the retention of entanglement at $k^\text{th}$ round. 
Now we choose the sharpness parameter, $\lambda_{k+1}$,  associated with the POVM $\{O_j^{k+1}\}$ used by the $(k+1)^\text{th}$ pair to discriminate the post-measurement state $\rho^k_b$ to be equal to $g^k_2$, i.e.,
\begin{equation}
\begin{aligned}
   & \lambda_{k+1}=g^k_2 \\
&= \left(1 + \epsilon_k\right) \max \left\{
\dfrac{1 + \Tr[\sigma_3 \otimes \sigma_3 \, \rho_1^k]}{\Tr[\sigma_2 \otimes \sigma_2 \, \rho_1^k]},
\dfrac{1 + \Tr[\sigma_3 \otimes \sigma_3 \, \rho_2^k]}{\Tr[\sigma_2 \otimes \sigma_2 \, \rho_2^k]}
\right\}.
\end{aligned}
\label{myeq7}
\end{equation}
Note that for the existence of the witness operator $\mathbb{W}^k_b$, one must ensure that {$0 \leq \lambda_{k+1} \leq 1$}. Thus, we need to determine up to which value of $k$ this condition holds, so that the existence of the witness operator $\mathbb{W}^k_b$ is confirmed. In Appendix~\ref{A1}, we show that if $\lambda_1 \to 0$, then the sequence~\eqref{sq} ensures $\lambda_k \to 0$, implying that such a witness operator $\mathbb{W}^k_b$ exists for any $k$. This result indicates that an arbitrarily large number of pairs can witness entanglement using the operator $\mathbb{W}^k_b$. Consequently, the SSDSE protocol can be sustained over infinitely many rounds without fully destroying the entanglement of the initial states.
We would like to emphasize here that the witness operators, $\mathbb{W}_b^k$, are introduced to show only the existence of entanglement in the post-measurement states, $\rho^k_b$, and no measurements are actually made to detect entanglement.}

Since we have ensured that the states remain entangled at each round for a particular choice of the sharpness parameter, let us now move to the second part of the SSDSE protocol which involves verifying that the success probability at each round remains greater than $\frac{1}{2}$ for the protocol to retain an advantage over random guess. We carry out this analysis in the following subsection.

\subsection{Success probability of discrimination in each round}
\label{S3P}

{In order to obtain success probability better than $P_R$}, Alice$_k$ and Bob$_k$ uses the set of measurement operators, $\{A_i^k\}$ and $\{B_{j|i}^k\}$, respectively, to discriminate the states handed to them by the $(k-1)^{\text{th}}$ pair. Here also the pairs stick to the guessing strategy $p_\text{opt}(1|0,0)=p_\text{opt}(1|1,0)=p_\text{opt}(2|0,1)=p_\text{opt}(2|1,1)=1$, even though, in this case, it may not be optimal. Hence, the average success probability for the $k^{\text{th}}$ pair to identify the state correctly is given by:
\begin{equation}
P_{\text{suc}}^{k} = \Tr[(O_1^k + O_3^k) \rho_1^{k-1}] + \Tr[(O_2^k + O_4^k) \rho_2^{k-1}].\label{myeq3}
\end{equation}
For $k = 1$, considering $\rho^0_1=\rho_1$ and $\rho^0_2=\rho_2$, we find the success probability is
\begin{equation*}
   P_{\text{suc}}^1 = \frac{1}{2} + \frac{1}{4} \left[ \lambda_1 + \lambda_1^2 - (-1 + \lambda_1)\lambda_1 \cos(2\theta) \right]. 
\end{equation*}
It is easy to verify that $ P_{\text{suc}}^1 > \frac{1}{2}$ for all $\theta \in (0, \pi/2]$ and $\lambda_1 \in (0,1)$. Thus, the first pair can successfully discriminate the states in the ensemble $\eta_2$ with a probability greater than the classical limit.

The question now is whether the subsequent pairs can also discriminate the states with a probability greater than $P_R$, while sustaining entanglement. If yes, what is the maximum value of $k$ up to which the aforementioned task is possible? We show that the task can be performed for arbitrarily large values of $k$, as long as $k \in \mathbb{N}$.

{To begin with, we substitute the expression of $O_j^k$ as given in Eq.~\eqref{jalsaghor} (with $\lambda = \lambda_k$) and the recurrence expression of $\rho_b^k$ in terms of $\rho_b^{k-1}$ (see  Eq.~\eqref{rst}) into Eq. }\eqref{myeq3}, and obtain:  
$P_{\text{suc}}^k = \frac{1}{2} + \sum_{b=1}^2\sum_{l=1}^{3} P_l^b(k)$,  
where $b=1$ and 2 correspond to the two possible input states from the ensemble $\eta_2$. The expressions of $P_l^b$ are given below:
\begin{eqnarray*}
P_1^b(k) &=&(-1)^{b+1}\frac{\lambda_k}{4} \Tr \Big[ \left( \mathbb{I}_2 \otimes \sigma_3 + \lambda_k \sigma_3 \otimes \sigma_3 \right)\rho_b^{k-1} \Big]\\
P_2^b(k) &=& (-1)^{b+1}\frac{\lambda_k \cos{2\theta}}{4} \Tr \Big[ \left( \mathbb{I}_2 \otimes \sigma_3 - \lambda_k \sigma_3 \otimes \sigma_3 \right)\\&&\hspace{6cm} \rho_b^{k-1} \Big], \\
P_3^b(k)&=& (-1)^{b+1}\frac{\lambda_k \sin{2\theta}}{4} \Tr \Big[ \left( \mathbb{I}_2 \otimes \sigma_1 - \lambda_k \sigma_3 \otimes \sigma_1 \right) \\&&\hspace{6cm} \rho_b^{k-1} \Big].
\end{eqnarray*}

As given in Eqs. \eqref{cof}, the terms of the form $\Tr\big[\sigma_p \otimes \sigma_q \rho_b^{k-1}\big]$, with $p = 0,3$ and $q = 1,3$, 
can be expressed in terms of $\Tr\big[\sigma_p \otimes \sigma_q \rho_b\big]$, where
for notational convenience, we defined $\sigma_0 = \mathbb{I}_2$. Hence, using those expressions, $\sum_{l=1}^{3} P_l^b(k)$ can be simplified to
\begin{widetext}
\begin{align}
    Q_k^b=\sum_{l=1}^{3} P_l^b(k)=\overbrace{\dfrac{\lambda_k\mu_b}{4}\Big(R_1^{k-1}+R_2^{k-1}\Big)\Big[1+\lambda_k +\big(1-\lambda_k\big)\cos{2\theta}\Big]}^{\mathcal{A}_k^b}+\overbrace{\dfrac{\lambda_k\big(1-\lambda_k\big)\mu_b}{4}\sin{2\theta}\Big(R_3^{'k-1}+R_4^{'k-1}\Big)}^{\mathcal{B}_k^b}\nonumber\\+\overbrace{\frac{\lambda_k}{4}
   \sin{2\theta}(1-\mu_b)\Big(R_3^{k-1}-R_4^{k-1}\Big)\Big[1-\lambda_k +\big(1+\lambda_k\big)\cos{2\theta}\Big]}^{\mathcal{C}_k^b}+\overbrace{\frac{\lambda_k}{4}\sin^2{2\theta}(1-\mu_b)\big(1+\lambda_k\big)\Big(R_1^{'k-1}-R_2^{'k-1}\Big)}^{\mathcal{D}_k^b}\nonumber\\+\overbrace{\frac{\lambda_k}{4}
    \cos{2\theta}(1-\mu_b)\Big(R_1^{k-1}-R_2^{k-1}\Big)\Big[1-\lambda_k +\big(1+\lambda_k\big)\cos{2\theta}\Big]}^{\mathcal{E}_k^b}+\overbrace{\frac{\lambda_k}{4}\sin{2\theta}\cos{2\theta}(1-\mu_b)\big(1+\lambda_k\big)\Big(R_3^{'k-1}-R_4^{'k-1}\Big)}^{\mathcal{F}_k^b}.\label{myeq6}
\end{align}    
\end{widetext}
Thus, the goal now reduces to showing that, for each $b$, the quantity $Q_k^b$ remains positive for arbitrary values of $k$. We denote the six terms in the expression of $Q_k^b$ as $\mathcal{A}_k^b$, $\mathcal{B}_k^b$, $\mathcal{C}_k^b$, $\mathcal{D}_k^b$, $\mathcal{E}_k^b$, and $\mathcal{F}_k^b$ (as indicated in the expression of $Q^b_k$ in Eq. \eqref{myeq6}). Readers needs to be careful, that though the notation $\mathcal{A}_k^b$ looks similar with the notation, $A_i^k$, used to denote the unsharp measurement operators (see Eq. \eqref{myeq4}), they are not directly related.

To show that $Q_k^b > 0$ for all values of $\theta$, we divide the range of $\theta$ into two intervals. First, we consider $\theta \in \left(0, \frac{\pi}{4}\right]$. Note that for $0 < \theta \leq \frac{\pi}{4}$, we have:
\begin{itemize}
    \item $R_1^{k-1} \pm R_2^{k-1} > 0$,    
    \item $R_3^{\prime k-1} \pm R_4^{\prime k-1} > 0$,
    \item $R_3^{k-1} - R_4^{k-1} > 0$,
    \item $R_1^{\prime k-1} - R_2^{\prime k-1} > 0$.
\end{itemize}
Thus, it is straightforward to realize that $Q_k^b > 0$ for $\theta \in \left(0, \frac{\pi}{4}\right]$ and 
 for {$k\in \mathbb{N}$}. 

Next, we move our focus to the other interval of parameters, i.e., $\theta \in \left(\frac{\pi}{4}, \frac{\pi}{2}\right]$ and $\mu_b\in{(0,1)}$. 
Positivity of $Q_k^b > 0$ can also be proved within this range. The outline of the proof is given below (detailed proof can be found in Appendix \ref{S3B}): 
\begin{itemize}
    \item First we focus on the first two terms, i.e., $\mathcal{A}_k^b + \mathcal{B}_k^b$ and express them as
\begin{equation}
\mathcal{A}_k^b + \mathcal{B}_k^b =\dfrac{\lambda_k}{4} \mu_b f^{k-1}_{\lambda_k}(\theta),    
\end{equation}
where $f^{k-1}_{\lambda_k}(\theta)$ is a function of all $\lambda_{\tilde{k}}$ such that $\tilde{k} \leq k$, and the subscript $k-1$ indicates that the expression for $f^{k-1}_{\lambda_k}(\theta)$ contains the terms $\left(R_1^{k-1} + R_2^{k-1}\right)$ and $\left(R_3^{'k-1} + R_4^{'k-1}\right)$. First we numerically find that $\mathcal{A}_k^b + \mathcal{B}_k^b>0$ for $k=2$ in the the range $\theta \in \left(\frac{\pi}{4}, \frac{\pi}{2}\right]$ and $\mu_b\in{(0,1)}$. Then we analytically show $\mathcal{A}_k^b + \mathcal{B}_k^b>0$ for any finite integer $k\geq 3$ within the considered ranges of $\theta$ and $\mu_b$.
\item Then, similar to the case of the first two terms, we first numerically show that $\mathcal{C}_2^b + \mathcal{D}_2^b>0$, and then using this inequality, we find analytically that $\mathcal{C}_k^b + \mathcal{D}_k^b>0$ for any finite positive integer, $k$. 
\item 
Finally, we move our attention to $\mathcal{E}_k^b + \mathcal{F}_k^b$. We realize, $\mathcal{E}_2^b + \mathcal{F}_2^b$ can both be positive and negative. {However}, we show, $\mathcal{E}_2^b + \mathcal{F}_2^b>0$ implies $\mathcal{E}_k^b + \mathcal{F}_k^b>0$ for any arbitrary, finite, positive integer, $k$, with the help of numerical techniques.
Hence, up to now, we realize that though the function $\mathcal{E}_2^b + \mathcal{F}_2^b$, is not always positive, when it is, that implies $Q_k^b>0$. Let us now focus on the other case, i.e., for which {$\mathcal{E}_2^b + \mathcal{F}_2^b\leq 0$}.
    \item Numerically, we find that, even if $\mathcal{E}_2^b + \mathcal{F}_2^b\leq 0$, $\mathcal{E}_2^b + \mathcal{F}_2^b+\mathcal{C}_2^b + \mathcal{D}_2^b$ is always positive, further implying $Q_k^b>0$, for the entire range of parameters.
     \item
    We know, even if $\mathcal{E}_2^b + \mathcal{F}_2^b\leq 0$, both $\mathcal{C}_k^b+\mathcal{D}_k^b$ and $\mathcal{F}_k^b$ are positive {
    for arbitrary $k\geq 2$}. Using these properties, we show that if the set of the sharpness parameters, $\{\lambda_k\}$, is an increasing series of $k$, then $\mathcal{E}_k^b + \mathcal{F}_k^b+\mathcal{C}_k^b+\mathcal{D}_k^b>0$ for $k>2$. This completes our proof for the entire considered ranges of $\theta$ and $\mu_b$. We would like to mention here that to prove also the previous statements, {we have considered the series $\{\lambda_k\}$ as increasing with $k$.}
\end{itemize}
Thus, we have $Q_k^b > 0$ for all positive finite integer values of $k$, $0<\theta\leq\pi/2$, and $0<\mu_b<1$. Hence we can finally conclude our analysis by stating that one can always choose an increasing sequence of $\lambda_k$ values in the limit $\lambda_k \to 0$ for all $k$, such that the discrimination probability at each round of the SSDSE is always greater than $\frac{1}{2}$, irrespective of the value of $k$ as long as $k \in \mathbb{N}$. Therefore, the analysis results in the following theorem:
{\begin{theorem}
   There always exists unsharp measurements, using which the states, $\ket{\Phi_1} = \sqrt{\mu_1} \ket{00}_{AB} + \sqrt{1-\mu_1} \ket{1\varsigma_1}_{AB}$ and 
   $
    \ket{\Phi_2} = \sqrt{\mu_2} \ket{01}_{AB} + \sqrt{1-\mu_2} \ket{1\varsigma_1^\perp}_{AB}$, prepared with equal probabilities, can be sequentially distinguished, an arbitrary number of times, with probability more than $\frac{1}{2}$ (i.e., the probability of success by just randomly guessing the state), sustaining non-zero amount of entanglement in the post-measurement states, where $\ket{\varsigma_{1}}_B= \cos{\theta} \ket{0}_B + \sin{\theta} \ket{1}_B$ and $\ket{\varsigma_{1}^\perp}_B = -\sin{\theta} \ket{0}_B + \cos{\theta} \ket{1}_B$, with $\mu_{1,2}\in (0,1)$ and $\theta\in (0, \pi/2]$. 
\end{theorem}}



\section*{Conclusion}
\label{con}
Entanglement plays a pivotal role in quantum communication tasks, serving as a key resource that enables functionalities such as device-independent secure key distribution, quantum teleportation, and quantum dense coding.
Distinguishing quantum states is also a fundamental task, necessary in several protocols, like entanglement distillation. 
However, conventional  state discrimination and sequential state strategies often lead to the destruction of entanglement, thereby limiting their usefulness in extended quantum protocols.

To address this limitation, we introduced a  protocol - sequential state discrimination while sustaining entanglement - that not only enabled unlimited sequential discrimination of states but also sustained a finite amount of entanglement in the output states.
{We initially applied our protocol to discriminate between two  orthogonal, entangled, two-qubit pure states from a large family of such couples prepared with equal prior probabilities. For such ensembles, we computed the exact success probability  and demonstrated that it consistently exceeded the probability achievable via random guessing for an arbitrary number of steps. Indeed, the success probability can be made arbitrarily close to unity, while retaining entanglement of the states at each step. We  evaluated the entanglement of the resulting states at each step, using logarithmic negativity as the entanglement measure, showing that the entanglement remained finite (i.e., non-zero) regardless of the number of pairs participating in the discrimination sequence.}

{Subsequently, we extended our protocol to an ensemble comprising of two arbitrary orthogonal, entangled, two-qubit pure states with equal prior probabilities. For this scenario, we established that it is always possible to construct a witness operator capable of verifying the presence of entanglement in the states at the end of each round of the sequential discrimination process, while ensuring that the success probability at each step exceeds that of random guessing.}

{Our protocol, thus, enables the transmission of classical information through local state discrimination to an arbitrary number of pairs of parties while ensuring that the state at the end of each step remains resourceful (i.e., entangled). Consequently, any pair that chooses not to decode the complete information can still utilize the entangled states for other quantum communication tasks where entanglement serves as a key resource. Also, our proposed strategy guarantees successful sequential local discrimination of the two arbitrary pure orthogonal entangled states by an arbitrary number of parties, with a success probability always exceeding that of random guessing. This potentially paves the way for more robust quantum information processing by balancing successful local state discrimination with long-term entanglement retention.}

\section*{Acknowledgment}
KS acknowledges support from the project MadQ-CM (Madrid Quantum de la Comunidad de Madrid) funded by the European Union (NextGenerationEU, PRTR-C17.I1) and by the Comunidad de Madrid (Programa de Acciones Complementarias).



\onecolumngrid
\appendix
\section{Recursion relation for states obtained 
the observer pairs in the SSDSE}
\label{kya bolti h public}
Since the post-measurement states at $k^\text{th}$ round can always be written in terms of the post-measurement state at  $(k-1)^{th}$ round as,
\begin{equation*}
\rho^{k}_{b}=\sum_{j=1}^4\sqrt{O^k_j}\rho_b^{k-1}\sqrt{O^k_j},
\end{equation*}
where $i=1,2$ corresponds to the first and second state of the ensemble $\eta_2$, respectively, so to get the explicit form of the post measurement state $\rho_{b}^k$, at $k^\text{th}$ round, we need to know the forms of $\sqrt{O^k_j}$ for each $j = \{1,2,3,4\}$ and for all $k \in \mathbb{N}$.
Note that,
\begin{eqnarray*}     \sqrt{O_1^{k}}=\sqrt{A_0^{k}}\otimes \sqrt{B_{0|0}^k}, \hspace{5 mm} \sqrt{O_2^{k}}=\sqrt{A_0^{k}}\otimes \sqrt{B_{1|0}^k}, \hspace{5 mm} \sqrt{O_3^{k}}=\sqrt{A_1^{k}}\otimes \sqrt{B_{0|1}^k}, \hspace{5 mm} 
\sqrt{O_4^{k}}=\sqrt{A_1^{k}}\otimes \sqrt{B_{1|1}^k}.
\end{eqnarray*}

Since any operator in $\mathbb{C}^2$ can be decomposed in the basis $\{\sigma_q\}$, where $q =1, 2, 3$, we can write $\sqrt{A_m^{k}}$ and $\sqrt{B_m^{lk}}$ (with $m\in \{0,1\}$, $j\in \{1,2\}$ and $k \in \mathbb{N}$ as follows:
\begin{align*}
&\sqrt{A_0^{k}}=\sqrt{B_{0|0}^k}=
\frac{1}{2\sqrt{2}}\Bigg[\Big(\sqrt{1+\lambda_k}+\sqrt{1-\lambda_k}\Big)\mathbb{I}_2
+\Big(\sqrt{1+\lambda_k}-\sqrt{1-\lambda_k}\Big)\sigma_3\Bigg],\\
&\sqrt{A_1^{k}}=\sqrt{B_{1|0}^k}=
\frac{1}{2\sqrt{2}}\Bigg[\Big(\sqrt{1+\lambda_k}+\sqrt{1-\lambda_k}\Big)\mathbb{I}_2
-\Big(\sqrt{1+\lambda_k}-\sqrt{1-\lambda_k}\Big)\sigma_3\Bigg], \\
&\sqrt{B_{0|1}^k}=
\frac{1}{2\sqrt{2}}\Bigg[\Big(\sqrt{1+\lambda_k}+\sqrt{1-\lambda_k}\Big)\mathbb{I}_2
+\Big(\sqrt{1+\lambda_k}-\sqrt{1-\lambda_k}\Big)\sigma'\Bigg],\\
&\sqrt{B_{1|1}^k}=\frac{1}{2\sqrt{2}}\Bigg[\Big(\sqrt{1+\lambda_k}+\sqrt{1-\lambda_k}\Big)\mathbb{I}_2
-\Big(\sqrt{1+\lambda_k}-\sqrt{1-\lambda_k}\Big)\sigma'\Bigg],
\end{align*}
where
$\sigma' \coloneqq \sigma_1\sin{2\theta}+\sigma_3\cos{2\theta}$.

Substituting the forms of $\sqrt{O^k_j}$ in Eq.~\eqref{st}, the post measured state at $k^\text{th}$ round, $\rho_{b}^k$, can be expressed in terms of the post measured state at $(k-1)^{th}$ round, $\rho_{i}^{k-1}$, as follows:

\begin{equation}
\begin{aligned}
\rho_{b}^k = \frac{1}{8} \Big[ &
2(1+\alpha_k)^2 \rho_{b}^{k-1}
+ (1+\alpha_k)(1-\alpha_k)\left(\mathbb{I}_2 \otimes \sigma_3 \rho_{b}^{k-1} \mathbb{I}_2 \otimes \sigma_3 + \mathbb{I}_2 \otimes \sigma' \rho_{b}^{k-1} \mathbb{I}_2 \otimes \sigma'\right) \\
& + (1-\alpha_k)^2 \left( \sigma_3 \otimes \sigma_3 \rho_{b}^{k-1} \sigma_3 \otimes \sigma_3 + \sigma_3 \otimes \sigma' \rho_{b}^{k-1} \sigma_3 \otimes \sigma' \right) + 2(1+\alpha_k)(1-\alpha_k)\left( \sigma_3 \otimes \mathbb{I}_2 \rho_{b}^{k-1} \sigma_3 \otimes \mathbb{I}_2 \right) \\
& + \lambda_k(1-\alpha_k) \Big( 
\mathbb{I}_2 \otimes \sigma_3 \rho_{b}^{k-1} \sigma_3 \otimes \sigma_3
- \mathbb{I}_2 \otimes \sigma' \rho_{b}^{k-1} \sigma_3 \otimes \sigma' + \sigma_3 \otimes \sigma_3 \rho_{b}^{k-1} \mathbb{I}_2 \otimes \sigma_3
- \sigma_3 \otimes \sigma' \rho_{b}^{k-1} \mathbb{I}_2 \otimes \sigma'
\Big)
\Big],
\end{aligned}
\label{rst}
\end{equation}
where $i=\{1,2\}$, and $\alpha_k \coloneqq \sqrt{1-\lambda_k^2}$.

\section{Entanglement witnessing by arbitrarily many pairs}
\label{A1}
In this section, we show how a particular choice of the sharpness parameter by the first pair (corresponding to $k=1$) ensures that the states passed on to an arbitrary pair in the sequence always remain entangled. Consequently, a witness operator $\mathbb{W}^k_b$ of the form given in Eq.~\eqref{wit} can always be constructed to detect the entanglement of these states. 

Recall that the sequence of sharpness parameters given in Eq.~\eqref{myeq7}, derived using the form of $\mathbb{W}^k_b$ in Eq.~\eqref{wit}, requires the expectation values of $\sigma_3 \otimes \sigma_3$ and $\sigma_2 \otimes \sigma_2$ over the relevant states.
Using the form of $\rho^{b}_k$ as given in Eq.~\eqref{rst}, these expectation values can be computed as:
\begin{align*}
\Tr\left[\sigma_3 \otimes \sigma_3 \rho_b^k\right] &= \left\{ \frac{1+\alpha_k}{2} + \frac{1-\alpha_k}{4} \Big(1 + \cos{(4\theta)}\Big) \right\} \Tr\left[\sigma_3 \otimes \sigma_3 \rho_b^{k-1}\right]+ \frac{1-\alpha_k}{4} \sin{(4\theta)} \Tr\left[\sigma_3 \otimes \sigma_1 \rho_b^{k-1}\right] \\
&\quad + \frac{\lambda_k (1-\alpha_k)}{4} \Big(1 - \cos{(4\theta)}\Big) \Tr\left[\mathbb{I}_2 \otimes \sigma_3 \rho_b^{k-1}\right] - \frac{\lambda_k (1 - \alpha_k)}{4} \sin{(4\theta)} \Tr\left[\mathbb{I}_2 \otimes \sigma_1 \rho_b^{k-1}\right], \\
\Tr\left[\sigma_2 \otimes \sigma_2 \rho_b^k\right] &= \alpha_k^2 \Tr\left[\sigma_2 \otimes \sigma_2 \rho_b^{k-1}\right],
\end{align*}
where $\alpha_k = \sqrt{1 - \lambda_k^2}$, $i=1,2$ and $\theta$ corresponds to the state parameters as defined in Eqs.~\eqref{sabdhaner mar nei},~\eqref{megh na chaite jol}.

Let us focus on the first pair, i.e., $k=1$. For this, we must have:
\begin{equation}
\begin{aligned}
\lambda_{k+1} 
&= \left(1 + \epsilon_k\right) \max \left\{
\dfrac{1 + \Tr[\sigma_3 \otimes \sigma_3 \, \rho_1^k]}{\Tr[\sigma_2 \otimes \sigma_2 \, \rho_1^k]},
\dfrac{1 + \Tr[\sigma_3 \otimes \sigma_3 \, \rho_2^k]}{\Tr[\sigma_2 \otimes \sigma_2 \, \rho_2^k]}
\right\}.
\label{sq}
\end{aligned}
\end{equation}
If the first pair chooses the sharpness parameter such that $\lambda_1 \to 0$, i.e., $\alpha_1 \to 1$, then it is easy to check that $\Tr\left[\sigma_3 \otimes \sigma_3 \rho_b^{1}\right] \approx \Tr\left[\sigma_3 \otimes \sigma_3 \rho_b^0\right]$, where $\rho_b^0$ (for $b = 1, 2$) are the initial pure entangled states in $\eta_2$.

Any pure entangled state $\ket{\psi}$ in $\mathbb{C}^2 \otimes \mathbb{C}^2$ can, up to local unitary, be written as:
\begin{equation*}
\ket{\psi} = \sqrt{m} \ket{01} + \sqrt{1 - m} \ket{10},
\end{equation*}
where $m \in \left(0, \frac{1}{2}\right]$ is the Schmidt coefficient. The corresponding density matrix, in Hilbert-Schmidt decomposition, can be expressed as:
\begin{equation*}
\frac{1}{4} \left[\mathbb{I}_4 - \sigma_3 \otimes \sigma_3 + 2\sqrt{m(1 - m)} \left(\sigma_1 \otimes \sigma_1 + \sigma_2 \otimes \sigma_2\right)\right].
\end{equation*}
The two initial entangled pure states in $\eta_2$ can also be decomposed in a similar fashion. Let $m_1$ and $m_2$ be their respective Schmidt coefficients. Thus, we obtain $\Tr[\sigma_3 \otimes \sigma_3 \rho_b^0] = -1$ for both $b=1,2$.
Hence, whenever $\lambda_1 \to 0$, we have $\Tr[\sigma_3 \otimes \sigma_3 \rho_b^1] \approx -1$, which implies $\lambda_2 \to 0$. Repeating the same logic, we get $\Tr[\sigma_3 \otimes \sigma_3 \rho_b^2] \approx -1$, and thus $\lambda_3 \to 0$, and so on.
Therefore, if initially $\lambda_1 \to 0$, then $\lambda_k \to 0$ for any $k$, implying that infinitely many pairs can witness the entanglement. This confirms that the states passed to any arbitrary pair always retain some entanglement.
\section{Success probability of discrimination in  each round}
\label{S3B}
The recursion relation of the state, $\rho_b^k$, in the $k^\text{th}$ round in terms of $(k-1)^\text{th}$ is given in Eq.~\eqref{rst}.  Recursively using that expression from the equation, one can find that the terms of the form $\Tr\big[\sigma_p \otimes \sigma_q \rho_b^{k-1}\big]$, with $p = 0,3$ and $q = 1,3$, 
can be expressed in terms of $\Tr\big[\sigma_p \otimes \sigma_q \rho_b\big]$. 
For notational convenience, we defined $\sigma_0 = \mathbb{I}_2$. The resulting expressions are:
\begin{equation}
\begin{aligned}
\Tr[\sigma_3 \otimes \sigma_3 \rho_b^{k-1}] &= 
  R_1^{k-1} \Tr[\sigma_3 \otimes \sigma_3 \rho_b] 
+ R_2^{k-1} \Tr[\mathbb{I}_2 \otimes \sigma_3 \rho_b]+ R_3^{k-1} \Tr[\sigma_3 \otimes \sigma_1 \rho_b] 
+ R_4^{k-1} \Tr[\mathbb{I}_2 \otimes \sigma_1 \rho_b], \\
\Tr[\mathbb{I}_2 \otimes \sigma_3 \rho_b^{k-1}] &= 
  R_2^{k-1} \Tr[\sigma_3 \otimes \sigma_3 \rho_b] 
+ R_1^{k-1} \Tr[\mathbb{I}_2 \otimes \sigma_3 \rho_b] + R_4^{k-1} \Tr[\sigma_3 \otimes \sigma_1 \rho_b] 
+ R_3^{k-1} \Tr[\mathbb{I}_2 \otimes \sigma_1 \rho_b], \\
\Tr[\sigma_3 \otimes \sigma_1 \rho_b^{k-1}] &= 
  R_1'^{k-1} \Tr[\sigma_3 \otimes \sigma_1 \rho_b] 
+ R_2'^{k-1} \Tr[\mathbb{I}_2 \otimes \sigma_1 \rho_b]+ R_3'^{k-1} \Tr[\sigma_3 \otimes \sigma_3 \rho_b] 
+ R_4'^{k-1} \Tr[\mathbb{I}_2 \otimes \sigma_3 \rho_b], \\
\Tr[\mathbb{I}_2 \otimes \sigma_1 \rho_b^{k-1}] &= 
  R_2'^{k-1} \Tr[\sigma_3 \otimes \sigma_1 \rho_b] 
+ R_1'^{k-1} \Tr[\mathbb{I}_2 \otimes \sigma_1 \rho_b]+ R_4'^{k-1} \Tr[\sigma_3 \otimes \sigma_3 \rho_b] 
+ R_3'^{k-1} \Tr[\mathbb{I}_2 \otimes \sigma_3 \rho_b].
\label{cof}
\end{aligned}
\end{equation}
The coefficients $R_j^{k}$ and $R_j^{\prime k}$ with $j = 1, 2, 3, 4$ can be determined from the following recursion relations:
\small
\begin{equation*}
    R_1^{k}=\Big[\dfrac{1+\alpha_k}{2}+\dfrac{1-\alpha_k}{4}\Big(1+\cos{4\theta}\Big)\Big] R_1^{k-1}+\dfrac{1-\alpha_k}{4}\sin{4\theta}R_3^{'k-1}+\dfrac{\lambda_k(1-\alpha_k)}{4}\Big(1-\cos{4\theta}\Big)R_2^{k-1}-\dfrac{\lambda_k(1-\alpha_k)}{4}\Big(\sin{4\theta}\Big)R_4^{'k-1},
\end{equation*}
\begin{equation*}
    R_2^{k}=\Big[\dfrac{1+\alpha_k}{2}+\dfrac{1-\alpha_k}{4}\Big(1+\cos{4\theta}\Big)\Big] R_2^{k-1}+\dfrac{1-\alpha_k}{4}\sin{4\theta}R_4^{'k-1}+\dfrac{\lambda_k(1-\alpha_k)}{4}\Big(1-\cos{4\theta}\Big)R_1^{k-1}-\dfrac{\lambda_k(1-\alpha_k)}{4}\Big(\sin{4\theta}\Big)R_3^{'k-1},
\end{equation*}
\begin{equation*}
    R_3^{k}=\Big[\dfrac{1+\alpha_k}{2}+\dfrac{1-\alpha_k}{4}\Big(1+\cos{4\theta}\Big)\Big] R_3^{k-1}+\dfrac{1-\alpha_k}{4}\sin{4\theta}R_1^{'k-1}+\dfrac{\lambda_k(1-\alpha_k)}{4}\Big(1-\cos{4\theta}\Big)R_4^{k-1}-\dfrac{\lambda_k(1-\alpha_k)}{4}\Big(\sin{4\theta}\Big)R_2^{'k-1},
\end{equation*}
\begin{equation*}
    R_4^{k}=\Big[\dfrac{1+\alpha_k}{2}+\dfrac{1-\alpha_k}{4}\Big(1+\cos{4\theta}\Big)\Big] R_4^{k-1}+\dfrac{1-\alpha_k}{4}\sin{4\theta}R_2^{'k-1}+\dfrac{\lambda_k(1-\alpha_k)}{4}\Big(1-\cos{4\theta}\Big)R_3^{k-1}-\dfrac{\lambda_k(1-\alpha_k)}{4}\Big(\sin{4\theta}\Big)R_1^{'k-1},
\end{equation*}
\begin{equation*}
    R_1^{'k}=\Big[\dfrac{1+\alpha_k}{2}+\dfrac{1-\alpha_k}{4}\Big(1+\cos{4\theta}\Big)\Big] R_1^{'k-1}+\dfrac{1-\alpha_k}{4}\sin{4\theta}R_3^{k-1}-\dfrac{\lambda_k(1-\alpha_k)}{4}\Big(1-\cos{4\theta}\Big)R_2^{'k-1}-\dfrac{\lambda_k(1-\alpha_k)}{4}\Big(\sin{4\theta}\Big)R_4^{k-1},
\end{equation*}
\begin{equation*}
    R_2^{'k}=\Big[\dfrac{1+\alpha_k}{2}+\dfrac{1-\alpha_k}{4}\Big(1+\cos{4\theta}\Big)\Big] R_2^{'k-1}+\dfrac{1-\alpha_k}{4}\sin{4\theta}R_4^{k-1}-\dfrac{\lambda_k(1-\alpha_k)}{4}\Big(1-\cos{4\theta}\Big)R_1^{'k-1}-\dfrac{\lambda_k(1-\alpha_k)}{4}\Big(\sin{4\theta}\Big)R_3^{k-1},
\end{equation*}
\begin{equation*}
    R_3^{'k}=\Big[\dfrac{1+\alpha_k}{2}+\dfrac{1-\alpha_k}{4}\Big(1+\cos{4\theta}\Big)\Big] R_3^{'k-1}+\dfrac{1-\alpha_k}{4}\sin{4\theta}R_1^{k-1}-\dfrac{\lambda_k(1-\alpha_k)}{4}\Big(1-\cos{4\theta}\Big)R_4^{'k-1}-\dfrac{\lambda_k(1-\alpha_k)}{4}\Big(\sin{4\theta}\Big)R_2^{k-1},
\end{equation*}
\begin{equation*}
    R_4^{'k}=\Big[\dfrac{1+\alpha_k}{2}+\dfrac{1-\alpha_k}{4}\Big(1+\cos{4\theta}\Big)\Big] R_4^{'k-1}+\dfrac{1-\alpha_k}{4}\sin{4\theta}R_2^{k-1}-\dfrac{\lambda_k(1-\alpha_k)}{4}\Big(1-\cos{4\theta}\Big)R_3^{'k-1}-\dfrac{\lambda_k(1-\alpha_k)}{4}\Big(\sin{4\theta}\Big)R_1^{k-1}.
\end{equation*}
Using the above set of equations, $\sum_{l=1}^{3} P_l^b(k)$ can be simplified to

\begin{align*}
    Q_k^b=\sum_{l=1}^{3} P_l^b(k)&=\dfrac{\lambda_k}{4}\Bigg[\mu_b\Big[\Big(R_1^{k-1}+R_2^{k-1}\Big)\Big(1+\lambda_k +\big(1-\lambda_k\big)\cos{2\theta}\Big)+\big(1-\lambda_k\big)\sin{2\theta}\Big(R_3^{'k-1}+R_4^{'k-1}\Big)\Big]\\&+
   \sin{2\theta}(1-\mu_b)\Big[\Big(R_3^{k-1}-R_4^{k-1}\Big)\Big(1-\lambda_k +\big(1+\lambda_k\big)\cos{2\theta}\Big)+\big(1+\lambda_k\big)\sin{2\theta}\Big(R_1^{'k-1}-R_2^{'k-1}\Big)\Big]\\&+
    \cos{2\theta}(1-\mu_b)\Big[\Big(R_1^{k-1}-R_2^{k-1}\Big)\Big(1-\lambda_k +\big(1+\lambda_k\big)\cos{2\theta}\Big)+\big(1+\lambda_k\big)\sin{2\theta}\Big(R_3^{'k-1}-R_4^{'k-1}\Big)\Big]\Bigg].
\end{align*}    
Thus, the goal now reduces to showing that, for each $b$, the quantity $Q_k^b$ remains positive for arbitrary values of $k$. We denote the six terms in the expression of $Q_k^b$ as $\mathcal{A}_k^b$, $\mathcal{B}_k^b$, $\mathcal{C}_k^b$, $\mathcal{D}_k^b$, $\mathcal{E}_k^b$, and $\mathcal{F}_k^b$. Readers needs to be careful, that though the notation $\mathcal{A}_k^b$ looks similar with the notation, $A_i^k$, used to denote the unsharp measurement operators (see Eq. \eqref{myeq4}), they are not directly related.

To show that $Q_k^b > 0$ for all values of $\theta$, we first divide the range of $\theta$ into two intervals. First, we consider $\theta \in \left(0, \frac{\pi}{4}\right]$. Note that for $0 < \theta \leq \frac{\pi}{4}$, we have:
\begin{itemize}
    \item $R_1^{k-1} \pm R_2^{k-1} > 0$,    
    \item $R_3^{\prime k-1} \pm R_4^{\prime k-1} > 0$,
    \item $R_3^{k-1} - R_4^{k-1} > 0$,
    \item $R_1^{\prime k-1} - R_2^{\prime k-1} > 0$.
\end{itemize}

Thus, it is straightforward to verify that $Q_k^i > 0$ for $\theta \in \left(0, \frac{\pi}{4}\right]$. 

Next, we consider the case $\theta \in \left(\frac{\pi}{4}, \frac{\pi}{2}\right]$. From this point onward, our analysis will focus exclusively on this range of $\theta$. First we show the positivity of $\mathcal{A}_k^b + \mathcal{B}_k^b$. We can express $\mathcal{A}_k^b + \mathcal{B}_k^b$ as
\small
\begin{align*}
\mathcal{A}_k^b + \mathcal{B}_k^b &= \dfrac{\lambda_k}{4} \mu_b \Big[ \left(R_1^{k-1} + R_2^{k-1}\right)\Big(1 + \lambda_k +(1 - \lambda_k)\cos{2\theta} \Big)  + \Big(1 - \lambda_k\Big)\sin{2\theta}\left(R_3^{'k-1} + R_4^{'k-1}\right) \Big] \\
&= \dfrac{\lambda_k}{4} \mu_b f^{k-1}_{\lambda_k}(\theta).
\end{align*}
\begin{figure}
\includegraphics[scale=0.7]{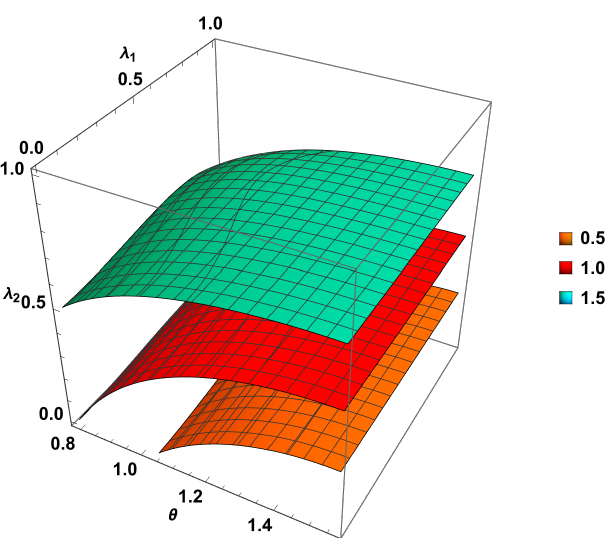}
\caption{
    Contour plot illustrating three positive surfaces of the function $f_{\lambda_2}^{1}(\theta)$  over the parameter range $\theta \in \left( \frac{\pi}{4}, \frac{\pi}{2} \right]$ and $\lambda_{1,2}\in(0,1)$. 
    By manipulating the parameter ranges, it is found that $f_{\lambda_2}^{1}(\theta)$ remains positive throughout the concerned range. 
    This figure depicts three such positive surfaces: the green surface corresponds to $f_{\lambda_2}^{1}(\theta) = 1.5$, the red surface represents $f_{\lambda_2}^{1}(\theta) = 1.0$, and the orange surface corresponds to $f_{\lambda_2}^{1}(\theta) = \frac{1}{2}$. 
    This plot serves as an example demonstrating that $f_{\lambda_2}^{1}(\theta)$ remains positive over the parameter range. The vertical axis is dimensionless, while the horizontal axes are dimensionless for $\lambda_1$ and in radians for $\theta$.
}
\label{f2}
\end{figure}
\normalsize
Here, $f^{k-1}_{\lambda_k}(\theta)$ is a function of all $\lambda_{\tilde{k}}$ such that $\tilde{k} \leq k$, and the superscript $k-1$ indicates that the expression for $f^{k-1}_{\lambda_k}(\theta)$ contains the terms $\left(R_1^{k-1} + R_2^{k-1}\right)$ and $\left(R_3^{'k-1} + R_4^{'k-1}\right)$.

For $k = 2$, we have $\dfrac{\lambda_2}{4}\mu_b > 0$, and the function $f^{1}_{\lambda_2}(\theta)$ depends on the parameters $\lambda_1$, $\lambda_2$, and $\theta$. Graphical analysis over the domain $\theta \in \left(\frac{\pi}{4}, \frac{\pi}{2}\right]$ with $\lambda_1, \lambda_2 \in (0,1)$ reveals that $f^{1}_{\lambda_2}(\theta)$ remains strictly positive throughout this range.
To illustrate this, in Fig.~\ref{f2}, we plot the values of $\lambda_2$ and $\theta$ for which $f^{1}_{\lambda_2}(\theta)=\frac{1}{2}$ (orange surface), $1.0$ (red surface), and $1.5$ (green surface). These three surfaces are presented as an exemplification, but it is important to note that additional positive surfaces can also be obtained by varying the parameters within the given ranges. 
Thus, our numerical analysis confirms that we have $\mathcal{A}^b_2 + \mathcal{B}^b_2 > 0$, with these three plotted surfaces serving as representative examples.

Next, we consider subsequent rounds with $k \geq 3$. We begin by noting that for the specified range of $\theta$, the following conditions hold:

\begin{itemize}
    \item $R_3^{'k} + R_4^{'k} < 0$
    \item $R_1^k + R_2^k > 0$
    \item The function $f^{k}_{\lambda_k}(\theta)$ satisfies:
    $
    f^{k}_{\lambda_k}(\theta) > 2\lambda_k(1 + \lambda_k)\left(1 - \sqrt{1 - \lambda_k^2}\right)\sin^2{\theta} \left(R_1^{k-1} + R_2^{k-1}\right) > 0.
    $
\end{itemize}
$f^{k-1}_{\lambda_k}(\theta) > 0$. Using the above conditions, we deduce that:
\begin{equation}
    f^{k-1}_{\lambda_k}(\theta) > \dfrac{2(\lambda_k - \lambda_{k-1})}{1 - \lambda_{k-1}} \left(R_1^{k-1} + R_2^{k-1}\right).
\end{equation}
Clearly, $f^{k-1}_{\lambda_k}(\theta) > 0$ if $\lambda_k > \lambda_{k-1}$, i.e., if $\lambda_k$ is an increasing sequence.
Recall the sequence of $\lambda_{k+1}$ defined in Eq.~\eqref{sq}. For this sequence to be increasing, we require:
\begin{equation*}
    \epsilon_{k+1} > \dfrac{T_k (1 + \epsilon_k)}{T_{k+1}} - 1,
\end{equation*}
where
{\begin{equation*}
    T_k = \max \left\{
    \dfrac{1 + \Tr[\sigma_3 \otimes \sigma_3 \, \rho_1^k]}{\Tr[\sigma_2 \otimes \sigma_2 \, \rho_1^k]},
    \dfrac{1 + \Tr[\sigma_3 \otimes \sigma_3 \, \rho_2^k]}{\Tr[\sigma_2 \otimes \sigma_2 \, \rho_2^k]}
    \right\}.
\end{equation*}}

Since the choice of $\epsilon^b_{k+1}$ is within our control, we can always select $\epsilon^b_{k+1}$ such that the inequality above is satisfied. 
Therefore, $\mathcal{A}_k^b + \mathcal{B}_k^b$ can be kept positive by choosing the sharpness parameters, ${\lambda_k}$, accordingly.

We now move on to the next component of the expression for $Q_k^b$, namely $\mathcal{C}_k^b + \mathcal{D}_k^b$. We can write:
\small
\begin{eqnarray*}
\mathcal{C}_k^b + \mathcal{D}_k^b 
&=& \dfrac{\lambda_k}{4} \sin{2\theta}(1 - \mu_b) \Big[ \left(R_3^{k-1} - R_4^{k-1}\right)\Big(1 - \lambda_k + (1 + \lambda_k)\cos{2\theta} \Big)  + (1 + \lambda_k)\sin{2\theta}\left(R_1^{'k-1} - R_2^{'k-1}\right) \Big] \\
&=& \dfrac{\lambda_k}{4} \sin{2\theta}(1 - \mu_b) \, g^{k-1}_{\lambda_k}(\theta).
\end{eqnarray*}
\normalsize
\begin{figure}
\includegraphics[scale=0.7]
{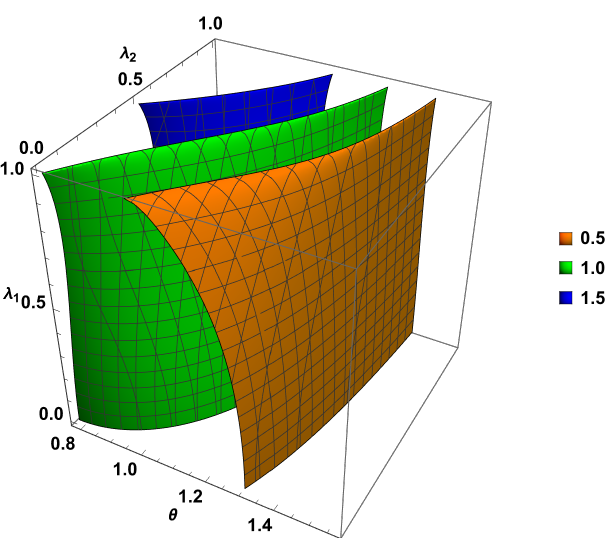}
\caption{
    Contour plot showing three positive surfaces of the function $g^{1}_{\lambda_2}(\theta)$ over the parameter range $\theta \in \left( \frac{\pi}{4}, \frac{\pi}{2} \right]$. This plot illustrates that $g^{1}_{\lambda_2}(\theta)$ remains strictly positive throughout the given range. 
    By varying the parameters, one consistently obtains surfaces with constant values of $g^{1}_{\lambda_2}(\theta)$.  As an exemplification of which we plot three surfaces. The blue surface corresponds to $g^{1}_{\lambda_2}(\theta) = 1.5$, the green surface to $g^{1}_{\lambda_2}(\theta) = 1.0$, and the orange surface to $g^{1}_{\lambda_2}(\theta) = \frac{1}{2}$. 
    The vertical axis is dimensionless, while the horizontal axes are dimensionless for $\lambda_2$ and measured in radians for $\theta$.
}

\label{f3}
\end{figure}
\normalsize
Similar as $f^{k-1}_{\lambda_k}(\theta)$, the function $g^{k-1}_{\lambda_k}(\theta)$ also depends on all $\lambda_{\tilde{k}}$ such that $\tilde{k} \leq k$. The subscript $k-1$ indicates that the expression for $g^{k-1}_{\lambda_k}(\theta)$ includes terms of the form $\left(R_3^{k-1} - R_4^{k-1}\right)$ and $\left(R_1^{'k-1} - R_2^{'k-1}\right)$. For the case $k = 2$, a numerical analysis over the relevant parameter range reveals that $g^{1}_{\lambda_2}(\theta) > 0$ throughout. To exemplify this behavior and support our argument, we have plotted three surfaces showing values of $\lambda_1$, $\lambda_2$, and $\theta$ for which $g^{1}_{\lambda_2}(\theta)=\frac{1}{2}$ (red surface), 1.0 (green surface), 1.5 (blue surface) in Fig.~\ref{f3}. Additional positive surfaces can similarly be obtained through further variation of the parameters. Since $\dfrac{\lambda_k}{4} \sin{2\theta} > 0$ for $\theta \in \left(\frac{\pi}{4}, \frac{\pi}{2}\right)$, it follows that $\mathcal{C}_2^b + \mathcal{D}_2^b > 0$.
 To show for $k\geq3$, $\mathcal{C}_k^b + \mathcal{D}_k^b>0$, we use
\begin{itemize}
    \item $\Big(R_1^{'k}-R_2^{'k}\Big)>0$,
    \item $\Big(R_3^{k}-R_4^{k}\Big)<0$.
    \label{c2}
\end{itemize}
For $k=3$, one can use these two conditions and the inequality
$C^i_2+D^i_2>0$, to find that 
\begin{equation*}
\begin{aligned}
\mathcal{C}_3^b + \mathcal{D}_3^b &>
\Bigg[\left(-1+\lambda_3- (1+\lambda_3) \cos{2\theta}\right) \Bigg(1+\sqrt{1-\lambda_2^2}+ 
   \dfrac{1}{2}\left(-1-\sqrt{1 - \lambda_2^2}\right)+ 
    \dfrac{1}{4}\left(1 - \sqrt{1 - \lambda_2^2}\right)(1- \lambda_2\\&+ (1+\lambda_2)\cos{4\theta})\Bigg)-\dfrac{1}{4}(1 + \lambda_3)(1 + \lambda_2) \left(1 - \sqrt{1 - \lambda_2^2}\right)\sin{2\theta}\sin{4\theta}+\dfrac{1}{1 + \lambda_2}
 \abs{1-\lambda_2 + (1 + \lambda_2)\cos{2 \theta}}\\&\csc{2\theta}\Bigg((1+\lambda_3)\left(\dfrac{1}{2}\left(1 + \sqrt{1 - \lambda_3^2}\right) - 
       \dfrac{1}{4}\left(1 - \sqrt{1 - \lambda_2^2}\right)(1- \lambda_2 + \left(1 + \lambda_2)\cos{4\theta}\right)\right) \sin{2\theta}+ 
     \dfrac{1}{4}(1 + \lambda_2)\\& \left(1 - \sqrt{1 - \lambda_2^2}\right) (1 - \lambda_3+ (1 + \lambda_3) \cos{2\theta})\sin{4\theta}\Bigg)\Bigg]\abs{\Big(R_3^{1}-R_4^{1}\Big)}\\ &= h(\lambda_3,\lambda_2,\theta)\abs{\Big(R_3^{1}-R_4^{1}\Big)}.
      \end{aligned}
\end{equation*}
Here, $h(\lambda_3,\lambda_2,\theta)$ is a function of $\lambda_3$, $\lambda_2$, and $\theta$. It can be shown that $h(\lambda_3,\lambda_2,\theta) > 0$ whenever $\lambda_3 > \lambda_2$ in the limit $\lambda_3$, $\lambda_2 \to 0$. This limit is also essential for the construction of a witness operator involving an infinite number of rounds, as discussed in the previous section.

In the same way, we can show if $\mathcal{C}_{k-1}^b + \mathcal{D}_{k-1}^b > 0$, then
\begin{equation*}
\mathcal{C}_k^b + \mathcal{D}_k^b > h(\lambda_k, \lambda_{k-1}, \theta) \left|R_3^{k-2} - R_4^{k-2}\right| > 0,
\end{equation*}
as long as $\lambda_k > \lambda_{k-1}$. Thus, it suffices that the sequence $\{\lambda_k\}$ is strictly increasing, which boils down to the same condition on $\epsilon_k^b$ that we found for $\mathcal{A}_k^b + \mathcal{B}_k^b$ to be positive for all $k$. 

\begin{figure}
\includegraphics[scale=0.7]
{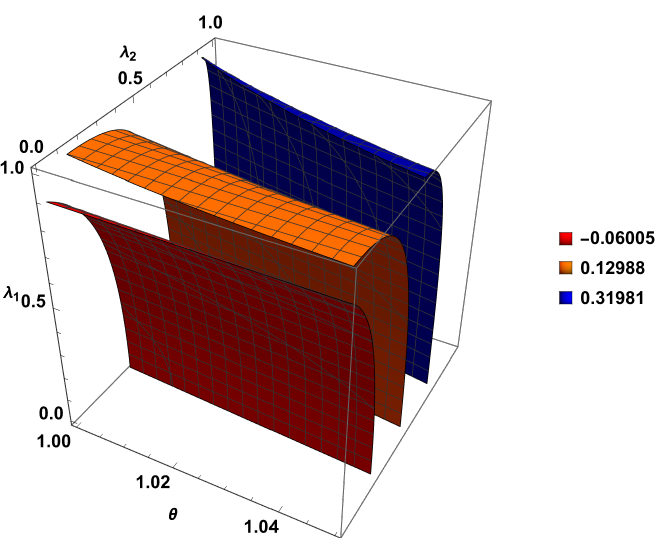}
\caption{Contour plot showing three representative constant-value surfaces of the function $t^{1}_{\lambda_2}(\theta)$. By varying the parameters $\lambda_1$, $\lambda_2$, and $\theta$, this function can attain both positive and negative values depending on the parameter choices. While many such surfaces can be obtained, we display three for illustration. The red surface corresponds to a constant negative value of $t^{1}_{\lambda_2}(\theta) = -0.0605$, while the blue and orange surfaces correspond to constant positive values of $0.12998$ and $0.31981$, respectively. The vertical axis is dimensionless; the horizontal axis for $\lambda_2$ is also dimensionless, and the axis for $\theta$ is in radians.}
\label{f4}
\end{figure}

Next, we turn to the final two terms, namely $\mathcal{E}_k^b + \mathcal{F}_k^b$. Note that one can write:
\begin{eqnarray*}
\mathcal{E}_k^b + \mathcal{F}_k^b 
&=& \dfrac{\lambda_k}{4}(1 - \mu_b) \Bigg[
\cos{2\theta} \Bigg(
\Big(R_1^{k-1} - R_2^{k-1}\Big) \Big(1- \lambda_k + (1 + \lambda_k)\cos{2\theta}\Big)  + (1 + \lambda_k)\sin{2\theta}\left(R_3^{'k-1} - R_4^{'k-1}\right)
\Bigg)
\Bigg] \\
&=& \dfrac{\lambda_k}{4}(1 - \mu_b)\, t^{k-1}_{\lambda_k}(\theta).
\end{eqnarray*}
Here also, the function, $t^{k-1}_{\lambda_k}(\theta)$, depends on all $\lambda_{\tilde{k}}$ for $\tilde{k} \leq k$, and the suffix $k-1$ indicates that the expression for $t^{k-1}_{\lambda_k}(\theta)$ includes the terms $(R_1^{k-1} - R_2^{k-1})$ and $(R_3^{'k-1} - R_4^{'k-1})$. As in earlier cases, we first consider $k = 2$. By numerically varying $\lambda_1$, $\lambda_2$, and $\theta$ we find $t^{1}_{\lambda_2}(\theta)$ can take both positive and negative values depending on the choice of parameters. As examples, notice Fig.~\ref{f4}, 
where we present surfaces representing values of $\lambda_1$, $\lambda_2$, and $\theta$ for which  $t^{1}_{\lambda_2}(\theta)$=-0.0605 (red surface), $0.12998$ (orange surface), and $0.31981$ (blue surface). Hence the red surface represents a subset of values of the parameters, $\lambda_1$, $\lambda_2$, and $\theta$, for which $t^{1}_{\lambda_2}(\theta)$ is negative. We emphasise that these three surfaces are selected examples, and other surfaces with both positive and negative values may also exist depending on the specific parameter choices. {If $\lambda_1$, $\lambda_2$, and $\theta$ take values such that $t^{1}_{\lambda_2}(\theta) > 0$ then $\mathcal{E}_2^b + \mathcal{F}_2^b$ would be positive implying $Q_2^b > 0$ and consequently  making $P_{suc}^k> \frac{1}{2}$ for $k = 2$.}

To examine whether $\mathcal{E}_k^b + \mathcal{F}_k^b > 0$ holds for arbitrary $k$, whenever $t^{1}_{\lambda_2}(\theta) > 0$, we begin by noting that for the considered parameter range $\theta \in \left(\frac{\pi}{4}, \frac{\pi}{2}\right]$, the following conditions always hold:
\begin{itemize}
    \item $(R_1^k - R_2^k) > 0$,
    \item $(R_3^{'k} - R_4^{'k}) < 0$.
\end{itemize}
Starting from $k = 3$, and using the conditions $(R_1^{k-2} - R_2^{k-2}) > 0$, $(R_3^{'k-2} - R_4^{'k-2}) < 0$, and $t^{k-2}_{\lambda_{k-1}}(\theta) > 0$, it can be numerically  verified that $t^{k-1}_{\lambda_k}(\theta) > 0$, provided that $\lambda_k > \lambda_{k-1}$. Therefore, for any increasing sequence $\{\lambda_k\}$, a positive value of $t^{1}_{\lambda_2}(\theta)$ ensures that $t^{k-1}_{\lambda_k}(\theta) > 0$ for arbitrary values of $k$. Consequently, when $t^{1}_{\lambda_2}(\theta) > 0$, one can always choose a suitable increasing sequence $\{\lambda_k\}$ such that $\mathcal{E}_k^b + \mathcal{F}_k^b > 0$, which implies $Q_k^b > 0$ and $P_{suc}^k>\frac{1}{2}$. As discussed earlier, such an increasing sequence of $\lambda_k$ can always be chosen without violating the requirement of. 

\begin{figure}
\includegraphics[scale=0.7]
{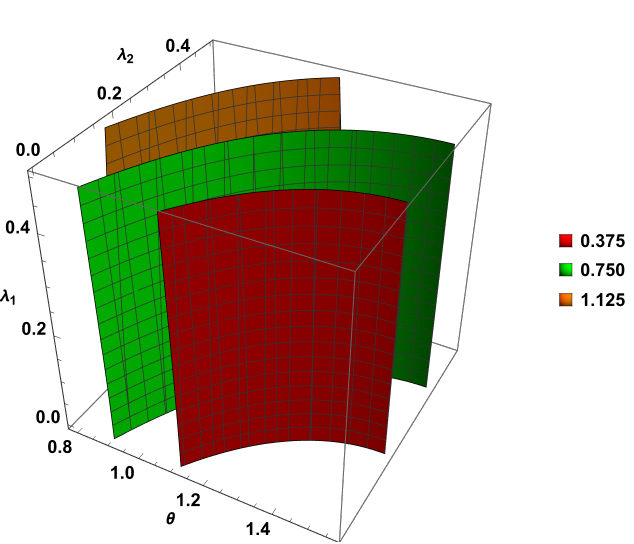}
\caption{Contour plot depicting three positive surfaces of the function $s^{1}_{\lambda_2}(\theta)$, as an exemplification over the parameter range. By adjusting the parameter values, one can find surfaces corresponding to constant positive values of $s^{1}_{\lambda_2}(\theta)$, suggesting that this function remains positive within the considered parameter range. In this plot, we present three such surfaces: $s^{1}_{\lambda_2}(\theta) = 0.375$ (red), $s^{1}_{\lambda_2}(\theta) = 0.750$ (green), and $s^{1}_{\lambda_2}(\theta) = 1.125$ (orange). The vertical axis is dimensionless, while the horizontal axes are dimensionless for $\lambda_2$ and expressed in radians for $\theta$.}
\label{f5}
\end{figure}

Next, we consider the case $t^{1}_{\lambda_2}(\theta) < 0$, and examine whether $Q_k^b$ can still be positive. For this case, we examine the positivity of $\mathcal{E}_k^b+\mathcal{F}_k^b+\mathcal{C}_k^b+\mathcal{D}_k^b$, which can be written as
\begin{equation*}
\begin{split}
\mathcal{E}_k^b+\mathcal{F}_k^b+\mathcal{C}_k^b+\mathcal{D}_k^b&= \dfrac{\lambda_k}{4}(1 - \mu_b)\left( t^{k-1}_{\lambda_k}+\sin{(2\theta) g^{k-1}_{\lambda_k}}\right)\\
&=\dfrac{\lambda_k}{4}(1 - \mu_b)s^{k-1}_{\lambda_k}(\theta).
\end{split}
\end{equation*}
As earlier, we first consider $k=2$
and numerically check that $s^{1}_{\lambda_2}(\theta)$ is always positive for the relevant ranges of $\theta$, $\lambda_1$, and $\lambda_2$. Plots illustrating three surfaces showing values of $\theta$, $\lambda_1$, and $\lambda_2$, for which $s^{1}_{\lambda_2}(\theta)$= 0.375 (red surface), 0.750 (green surface), and 1.125 (orange surface), is presented in Fig.~\ref{f5}. Therefore, for $k = 2$, we have $s^{1}_{\lambda_2}(\theta) > 0$, which implies $Q^b_2 > 0$ and $P_\text{suc}^2>\frac{1}{2}$.

For the subsequent round, we know
\small
\begin{align*}
&\mathcal{E}_k^b+\mathcal{F}_k^b=\dfrac{\lambda_k}{4}(1-\mu_b)\Big(p_{\lambda_k}r_{\lambda_{k-1}}+q_{\lambda_k}X_{\lambda_{k-1}}\Big)\Big(R_1^{k-2}-R_2^{k-2}\Big)+\Big(p_{\lambda_k}X_{\lambda_{k-1}}+q_{\lambda_k}Y_{\lambda_{k-1}}\Big)\Big(R_3^{'k-2}-R_4^{'k-2}\Big),
\end{align*}
where
\begin{align*}
&p_{\lambda_k} \coloneqq \cos{2\theta}\Big(1-\lambda_k+\big(1+\lambda_k\big)\cos{2\theta}\Big), \hspace{2 mm}\text{ }q_{\lambda_k} \coloneqq(1+\lambda_k)\sin{2\theta}\cos{2\theta}, \hspace{2 mm}X_{\lambda_k} \coloneqq \dfrac{(1-\alpha_k)(1+\lambda_k)}{4}\text{, }\alpha_k=\sqrt{1-\lambda_k^2},\\
&r_{\lambda_k} \coloneqq \Big[\dfrac{1+\alpha_k}{2}+\dfrac{1-\alpha_k}{4}\Big(1+\cos{4\theta}\Big)-\dfrac{\lambda_k(1-\alpha_k)}{4}\Big(1-\cos{4\theta}\Big)\Big],\\&
Y_{\lambda_k} \coloneqq \Big[\dfrac{1+\alpha_k}{2}-\dfrac{1-\alpha_k}{4}\Big(1+\cos{4\theta}\Big)-\dfrac{\lambda_k(1-\alpha_k)}{4}\Big(1-\cos{4\theta}\Big)\Big].
\end{align*}    
\normalsize
It can be checked numerically that $\big(p_{\lambda_k}X_{\lambda_{k-1}} + q_{\lambda_k}Y_{\lambda_{k-1}}\big) < 0$ and $q_{\lambda_k}X_{\lambda_{k-1}} > 0$ {within the range $\lambda_k \in(0,1) \forall  k$. Thus, the only term that can make $\mathcal{E}_k^b+\mathcal{F}_k^b$ negative is $p_{\lambda_k}r_{\lambda_{k-1}}$. If one plots $p_{\lambda_k}r_{\lambda_{k-1}}$ over the concerned parameter range, it can be seen that for certain parameter values, $p_{\lambda_k}r_{\lambda_{k-1}}$ is positive, while for others, $p_{\lambda_k}r_{\lambda_{k-1}} < 0$. Whenever $p_{\lambda_k}r_{\lambda_{k-1}} > 0$, we have $\mathcal{E}_k^b+\mathcal{F}_k^b> 0$. However, when $p_{\lambda_k}r_{\lambda_{k-1}} < 0$, we use the following conditions:}
\begin{itemize}
    \item $\Big(R_1^{k}-R_2^{k}\Big)>0$,
    \item $\Big(R_3^{'k}-R_4^{'k}\Big)<0$,
    \item $\mathcal{E}_k^b+\mathcal{C}_k^b+\mathcal{D}_k^b>0$,
    \item $\mathcal{C}_k^b+\mathcal{D}_k^b>0$,
\end{itemize}
that holds true for all $k$ and find 
\begin{align*}
    &\mathcal{E}_k^b+\mathcal{F}_k^b > a(\lambda_k,\lambda_{k-1},\theta)(R_1^{'k-2}-R_2^{'k-2})+c(\lambda_k,\lambda_{k-1},\theta)(R_3^{'k-2}-R_4^{'k-2}), 
 \end{align*}   
where
\begin{eqnarray*}
&&a(\lambda_k,\lambda_{k-1},\theta)= \Big(\dfrac{p_{\lambda_k}r_{\lambda_{k-1}}}{\abs{p_{\lambda_k}r_{\lambda_{k-1}}}}W_{\lambda_{k-1}}+ Z_{\lambda_{k-1}}X_{\lambda_{k}}+W_{\lambda_{k-1}}Y_{\lambda_k}\Big)\text{, and} \\
    &&c(\lambda_k,\lambda_{k-1},\theta)= \Big(\dfrac{p_{\lambda_k}r_{\lambda_{k-1}}}{\abs{p_{\lambda_k}r_{\lambda_{k-1}}}}Z_{\lambda_{k-1}}+ Z_{\lambda_{k-1}}r_{\lambda_{k}}+X_{\lambda_{k-1}}W_{\lambda_k}\Big),
\end{eqnarray*}   
with
\begin{align*}
    &W_{\lambda_{k}}=(1+\lambda_k)\sin^2{2\theta},\\
&Z_{\lambda_{k}}=\sin{2\theta}\Big(1-\lambda_k +\big(1+\lambda_k\big)\cos{2\theta}\Big).
\end{align*}
It can be shown numerically that if $\lambda_{k} > \lambda_{\tilde{k}}$ for $k > {\tilde{k}}$, i.e., if the sequence $\{\lambda_k\}$ is increasing, then $a(\lambda_k, \lambda_{k-1}, \theta)>0$ and $c(\lambda_k, \lambda_{k-1}, \theta)\big)<0$ in the limit $\lambda_k \to 0$ for all $k$. Thus, we have $Q_k^b > 0$ for all values of $k$. This completes our analysis, showing that one can always choose an increasing sequence of $\lambda_k$ values in the limit $\lambda_k \to 0$ for all $k$, such that the discrimination probability at each round of the SSDSE is always greater than $\frac{1}{2}$, irrespective of the value of $k$ as long as $k \in \mathbb{N}$. Moreover, such a choice also ensures that the relevant states retain entanglement for an arbitrarily large number of rounds.


In the next section, we turn to a specific example involving an ensemble of two well-defined entangled pure states, prepared with equal prior probability. By applying the SSDSE protocol to this ensemble, we compute the exact entanglement of the post-measurement states at each round, quantified via logarithmic negativity, and confirm that it remains nonzero throughout arbitrary number of rounds. Moreover, we compute the corresponding success probabilities, which consistently exceed the success probability of random guess, $\frac{1}{2}$.

\twocolumngrid

\bibliography{ref}
\end{document}